\documentclass{article}

\usepackage{arxiv}

\usepackage[utf8]{inputenc} 
\usepackage[T1]{fontenc}    
\usepackage{hyperref}       
\usepackage{url}            
\usepackage{booktabs}       
\usepackage{amsfonts}       
\usepackage{nicefrac}       
\usepackage{microtype}      
\usepackage{graphicx}
\usepackage{doi}

\usepackage[subrefformat=parens]{subcaption}
\usepackage{bm}
\usepackage{amsmath,amssymb}

\usepackage{here}

\title{Digital transformation of droplet/aerosol infection risk assessment realized on "Fugaku" for the fight against COVID-19}

\date{}

\author{ Kazuto Ando\thanks{All authors are listed in alphabetical order by surnames.} \\
	RIKEN Center for Computational Science, Japan\\
	\texttt{kazuto.ando@riken.jp} \\
	\And
	Rahul Bale$^*$ \\
	RIKEN Center for Computational Science, Japan\\
	\texttt{rahul.bale@riken.jp} \\
	\And
	ChungGang Li$^*$ \\
	RIKEN Center for Computational Science, Japan\\
	Kobe University, Japan \\
	\texttt{chung-gang.li@a.riken.jp} \\
	\And
	Satoshi Matsuoka$^*$ \\
	RIKEN Center for Computational Science, Japan\\
	Tokyo Institute of Technology, Japan \\
	\texttt{satoshi.matsuoka@riken.jp} \\
	\And
	Keiji Onishi$^*$ \\
	RIKEN Center for Computational Science, Japan \\
	\texttt{keiji.onishi@riken.jp} \\
	\And
	Makoto Tsubokura$^*$ \\
	RIKEN Center for Computational Science, Japan \\
	Kobe University, Japan \\
	\texttt{mtsubo@riken.jp} \\
}

\renewcommand{\headeright}{}
\renewcommand{\undertitle}{}
\renewcommand{\shorttitle}{}

\hypersetup{
pdftitle={Digital transformation of droplet/aerosol infection risk assessment realized on "Fugaku" for the fight against COVID-19},
pdfsubject={cs.CE, physics.flu-dyn},
pdfauthor={Kazuto Ando, Rahul Bale, ChungGang, Satoshi Matsuoka, Keiji Onishi, Makoto Tsubokura},
pdfkeywords={COVID-19, Computational fluid dynamics, Building cube method, Immersed boundary method, Dirty CAD, Droplet/Aerosol transmission, societal behavioral change},
}

\begin{document}
\maketitle

\begin{abstract}
The fastest supercomputer in 2020, Fugaku, has not only achieved digital transformation of epidemiology in allowing end-to-end, detailed quantitative modeling of COVID-19 transmissions for the first time, but also transformed the behavior of the entire Japanese public  through its detailed analysis of transmission risks in multitudes of societal situations entailing heavy risks. A novel aerosol simulation methodology was synthesized out of a combination of a new CFD methods meeting industrial demands, CUBE\cite{Jansson2019}, which not only allowed the simulations to scale massively with high resolution required for micrometer virus-containing aerosol particles, but also extremely rapid time-to-solution due to its ability to generate the digital twins representing multitudes of societal situations in minutes not week, attaining true overall application high performance; such simulations have been running for the past 1.5 years on Fugaku, cumulatively consuming top supercomputer-class resources and the result communicated by the media as well as becoming official public policies. 
\end{abstract}


%

\section{Introduction}

\subsection{COVID-19 Droplet/Aerosol Infection}

COVID-19, initially discovered in Wuhan, China at the end of 2019, quickly spread globally and changed our lives---however, the main question of this unknown virus was its main mode of transmission. In particular, in the early stages of the pandemic, various theories existed, some extrapolated from traditional epidemiological observations, but turning out to be somewhat incorrect, even in some cases, non-scientific, and sometimes such information coming from seemingly authoritative sources, significantly  disrupting our daily lives as well as socio-economic activities due to lockdown etc. One might still recall that, in the early stages of the pandemic, even institutions such as the World Health Organization (WHO) and the US Center for Disease Control and Prevention (CDC) gave skeptical announcement regarding the effectiveness of commercial surgical masks as preventive measures, which might have misdirected the behavior of people in the initial spread of the infection.

A month after WHO declared the COVID-19 pandemic, in March 2020 the Ministry of Education, Culture, Sports, Science and Technology (MEXT) and RIKEN Center for Computational Science (R-CCS), jointly announced to institute a program to rapidly exploit the computational capability of the new Fugaku supercomputer, which was still in the early days of its installation process, to combat COVID-19. As Fugaku would get gradually deployed, the COVID-19 applications would receive priority status with both resources and support from R-CCS and Fujitsu, the design and manufacturing partner of Fugaku; in the end, {\em each} of the project will be granted computing resources equalling the entire dominance of a top-tier class supercomputer, i.e., equivalent to tens of millions of node hours on a high-end machine, for duration of the year. Such a program was unprecedented, but was deemed necessary to counteract the global emergency. Overall, six projects were selected, and the program was conducted for a year, with some of the projects moving on to other schemes to effectively continue on Fugaku, including the aerosol/droplet simulations. 

In Japan, even before the global epidemic declaration, multiple cluster cases occurred in a large cruise ship, a sports gym, a large-scale festival venue, live music clubs etc., and epidemiological investigation quickly identified the main probable cause of transmission to be droplets and more predominantly aerosols, e.g., smaller droplets that are 5 microns or less, has been suggested. Unfortunately, precise epidemiology of such aerosol transmissions was not well developed, which we as many experts believed was the central cause of the confusion stated above. Indeed, various experiments attempting to capture the behavior of droplets and aerosols and the airflow affecting them had existed, but the experimental reproduction of end-to-end nature of the transmission was difficult at best, and almost impossible to apply to multitudes of societal conditions, involving multiple people in complex societal settings such as crowded trains, wearing various protective gears such as masks and face shields of different varieties. Moreover, the behavior of the aerosols being emitted from the lungs and other parts of the respiratory systems, as well as how they are captured by breathing them in through the nasal system to the lungs, was almost impossible to replicate via experiments, at least to produce some tangible quantitative results necessary to capture the risk parameters associated with various social situations such as social distancing. 

In order to accurately evaluate the risk of infection caused by droplets/aerosols and to propose effective measures to reduce the risk, it is necessary to accurately predict the total droplets volume reaching the surrounding individuals, after being emitted from an infected person with coughs, singing, shouting, etc., and transported by surrounding air in indoor environment. In addition, during  transportation, various physics need to considered such as droplets’ evaporation, adhesion to walls, and being lifted by thermal convection caused by heat being generated from human bodies, etc. Here, the obvious solution is to simulate such complex phenomena utilizing advanced multi-physics simulation using CFD as the basis. However, this was found to be non-trivial at best, as complex, moving boundaries has to be handled in extreme fine resolutions while the particle behavior must also be considered. Moreover, since the trajectory of droplets is greatly affected by the ambient airflow, , it is necessary to reproduce the operating conditions of various artifacts affecting airflow such as air ducts and air conditioning systems, together with the detailed geometry of target rooms, halls, vehicles, etc., as well as people situated inside e.g., crowded trains. This means that a huge number of simulations need to be performed for variety of indoor environments with complicated geometries in a limited amount of time, not only the simulation execution time, but also total runtime including mesh generation for complex shapes, which are often more of a bottleneck for the total time-to-solution. We believe such complexities were the exact reason that, only simple mathematical models are currently being used to model epidemiological transmissions, and detailed supercomputer simulation was not employed, but rather, restricted to experimental techniques (described in more detail below).

\begin{figure}[htb]
  \centering
    \includegraphics[keepaspectratio,width=\textwidth]{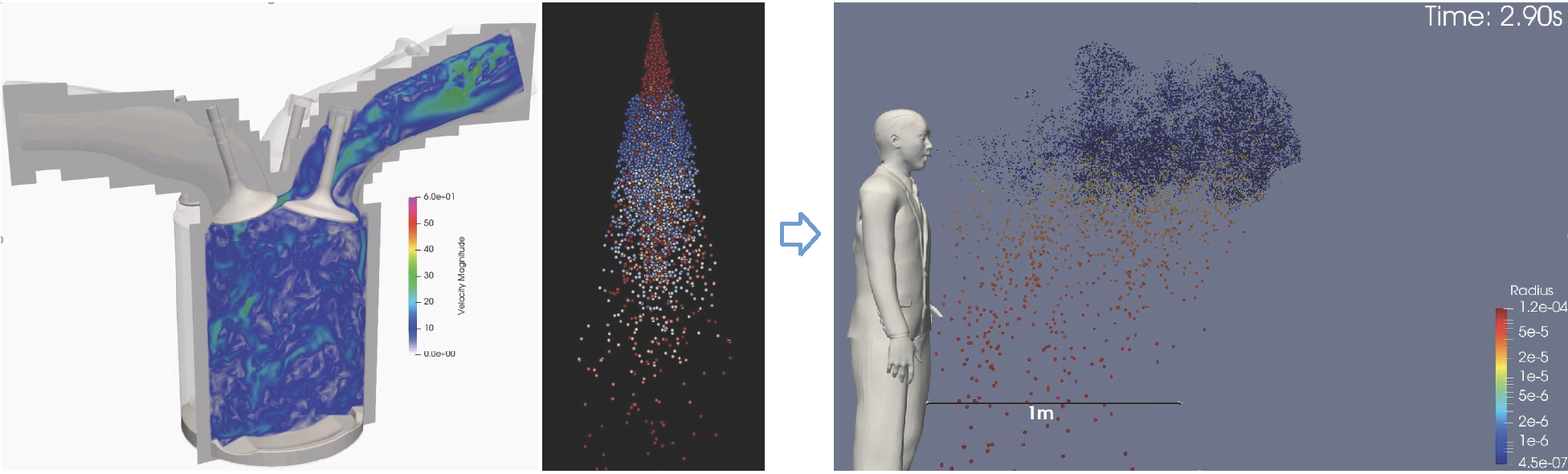}
   \caption{Fuel Injection of Internal Combustion Engine (left) vs. Droplet/Aerosol Spread (right)}
  \label{fig:injection}
\end{figure}

Fortunately, as we describe later, such a problem is very much synonymous to the problems facing advanced industrial design using supercomputers, in that thousands of design cases must be quickly evaluated precisely to reach an optimized  solution. In particular, advanced CFD techniques incorporated into an industrial-strength, massively-scalable engineering CFD program possible to handle geometries of billions of degrees of freedom as well as complex physics, was being developed as a co-design vehicle for Fugaku development, CUBE as mentioned, by a team being led by the last author, Makoto Tsubokura. Moreover, one of the target application of CUBE was the fuel injection of internal combustion engines, whose physics almost entirely matches that of aerosol emissions by humans\ref{fig:injection}. Such detailed simulations have never been performed to model droplets and aerosols, and the application of which, produced results never before realized with accurate physics, and its visualization changed the public behavior of the entire Japanese population almost overnight, once aired on all TV channels and spread through the Internet.

\section{Our Contributions Towards Mitigating COVID-19 Transmissions on Fugaku Supercomputer}

Our work best can be summarized as contributing towards COVID-19 transmission mitigation using supercomputers from three perspectives: {\em performance}, in that whole machine scalability and efficiency as well as orders of magnitude speedup in digital twin mesh generation was achieved, {\em technology}, in that a these being the result of the new CFD method combined with Fugaku/A64FX's HPC application-centric design, and {\em science and societal impact} in achieving transformational epidemiological simulation of droplets/aerosols and impacted the entire Japanese society's behavior to curtail the COVID-19 pandemic. More specifically:

\subsection{Technology} We have applied the novel CFD method and software CUBE\cite{Jansson2019}, which while being an efficient and scalable solver being based on explicit methods, but with a number of new provisions to allow it to be applied to real industrial applications much like the mainstream unstructured solvers, in fact being sufficiently accurate modeling of complex and difficult boundary conditions such as car window panes at extreme high resolutions required for modeling the detailed analysis of aerosols in real societal situations. Moreover, being based on explicit methods, mesh generations from 'dirty' CAD data can be effectively automated, in a matter of minutes instead of weeks typically required for unstructured models, allowing generations of multitudes of 'digital twins' representing accurate renditions of complex societal situations of classrooms, auditoriums, trains, busses, planes, restaurants etc. Finally driving the simulations is Fugaku, on which has been built with a novel A64FX processor with HBM2 integration for the first time as a general purpose processor, allowing CUBE to achieve 2x performance over other mainstream CPUs for HPC per node, and showing scalability up to nearly 100,000 nodes with high efficiency for a CFD solver.

\subsection{Performance} We have achieved high performance of the nodes as well as the ultra-high scalability of the simulation, as per traditional HPC metric, but we also achieved orders of magnitude performance improvements in the overall time-to-solution of the mesh generation, allowing ultra high throughput covering over 1000 societal situations of numerous varieties, more important than single-shot performance. Firstly, by the use of Fugaku and its A64FX processor our center has designed and created Good single node performance, being 2x\~3x over typical high performance CPU nodes used in top-class supercomputers. We also achieve high efficiency, weak scaling up to 50,000 nodes on Fugaku for simulation of a real vehicle model. Moreover, time to solution for creating 0.1mm meshes as required by droplet/aerosol simulations, with numerous {\em dirty} CAD data artifacts would require weeks to months according to our assessment, but rather, our methodologies would allow for a mesh in 10 minutes, achieving 3-4 orders of magnitude improvement. Overall, 1049 societal cases were simulated over the course of 17 months or approximately 62 cases per months throughput, including geometry generation from CAD data or laser point cloud measurements, actual droplet/aerosol simulation, and post-processing including visualization. The cumulative compute resource consumed would be almost equivalent to completely dedicating a top-tier supercomputer such as Univ. Tokyo Oakforest-PACS, which is former \#1 and currently the 5th fastest machine in Japan, during the same period.
     
\subsection{Science and Societal Impact} As far as we know, this is the first time detailed epidemiological capability for aerosol-borne infections with detailed end-to-end simulation capability has been realized on a leadership supercomputer, allowing detailed assessment of the risk factors involved in a variety of complicated societal transmission scenarios, whereas until now the field had been limited to physical experiments which were simply not possible to be conducted for such situations; as such we have achieved digital transformation for one of the most important scientific fields to combat COVID-19. Moreover, the results have had immense societal impact, achieving digital transformation of the entire Japanese population's behavior, with its results being broadcast on every TV station, newspapers, and social media in Japan, with the public, the government, and the industries reacting immediately. We believe we have contributed significantly to allow Japan to maintain one of the lowest infection rate among the G7 nations, despite constitutionally impossible to mandate lock-downs and thus the society kept comparably open, thus saving thousands of lives.

\section{Current State of the Art in Modeling of Viral Droplet/Aerosol Transmissions}

The current state of the art for precise, detailed simulations of droplet/aerosol transmissions on supercomputers had simply not existed as far as we know, and was the probable cause of numerous unfounded, seemingly unscientific statements by various parties as we mentioned earlier. We have uncovered several areas where the state of the art trails many other scientific disciplines, some being caused by the simplistic modeling of the phenomenon, and also, the sheer difficulty of modeling droplet/aerosol simulations, some of the innovations only realized by modern evolution of combustion modeling.

\subsection{Traditional Assessment of Infection Risks}
\label{subsect:infection}

The mainstream traditional methodology to numerically model the 
risk of infection/transmission by virus that has invaded a human body is the following simple formula.

\begin{equation}
	P(N)=1-e^{-\frac{N}{N_{0}}}
	\label{eq:infection_risk}
\end{equation}
Here, $N_{0}$ is the amount of virus that leads to infection, and $N$ is that consumed up by the human body. For airborne infectious diseases, The concept of instantaneous and uniform spread is commonly used to evaluate $N$, i.e., viral droplets/aerosols emitted by an infected person will immediately and uniformly spread within the surrounding room, and gradually dispersed by the ventilation system in the room. Typically, this is modeled as
\begin{equation}
	N=\frac{BS}{\lambda V}T\left\{1-\frac{1}{\lambda V}(1-e^{-\lambda T})\right\}
	\label{eq:dilution}
\end{equation}
Here, $B$ is the volume of breathed air by the person subject to infectious risk, S is the amount of virus emitted by the infected person, T is the exposure time, V is the room volume and lambda is the ventilation time in the room per hour. This risk evaluation is mainstream in epidemiology research, and in fact {\em is} effective for microbes and virus that are highly infectious such as tuberculosis and measles. On the other hand, for COVID-19, early research scrutinizing various infection scenarios suggested that inhaling high concentration of virus-ridden aerosol at a short distance from the infected person is the main path of transmission. Thus, the above evaluation is likely to underestimate the infection risk, and more precise prediction of spatio-temporal distribution of viral droplet/aerosol was obviously required for COVID-19. In particular, distribution of aerosol density around the person at risk would be strongly affected by the indoor air flow induced and affected by numerous environmental factors, such as fans and air-cons, mechanical ventilation systems, as well as geometrical feature in the room, not only the ones that exist naturally, but also risk-mitigation additions such as shields, and most importantly, the human bodies themselves, not just as geometry but also as heat sources causing convection flows. 

Due to the aerosols being very small in micrometers easily affected by such factors as well as being numerous, it became obvious that moving away from simple models one can calculate on a smart phone, and employing the most modern CFD simulation methods on supercomputers was absolutely necessary, which has not been tried before as far as we know, but why? Next, we describe the difficulty of applying traditional CFD methods and software packages for this application, due to scalability and inability to meet the time-to-solution requirement to address the COVID-19 pandemic crisis.

\subsection{Traditional Industry-Strength Engineering CFD using Unstructured Grids and their Shortcomings}

The CFD method currently used in the real industry emphasizes the reproducibility of complex geometries. Since the flow characteristics depends on the geometry, it becomes the main factor that determine product performance. Additionally, industrial products consist of complex structural members such as thin plates or pipes are assembled with sealing materials like as rubber or bolt, causing formation of extremely variant, detailed and complex flow paths. In HPC research on large machines we have excellent proposals of structured rectangular grids or hexahedral grids that facilitate uniform memory access and good scalability, but they are disadvantageous in reproducing complex shapes and achieving accuracy of the detailed flow paths, as required for real industry use, and as such are rarely put to commercial use, and for the reasons we describe, not readily applicable the droplet/aerosol simulations for COVID-19.

By all means, the converse is true for unstructured grids, in that (tetra, prism, pyramid, etc.) induce irregular data access and poor in performance and scaling, the latter often due to load balancing. Although there are there have been various proposals to improve efficiency, e.g., by changing from element loops to node loops, or by reconstructing arbitrary polyhedrons to reduce the number of elements, nonetheless it tends to be heavily bound by memory access, and difficult to load balance, even with techniques like AMR. 

\begin{figure}[H]
	\centering
	\includegraphics[width=0.3\textwidth]{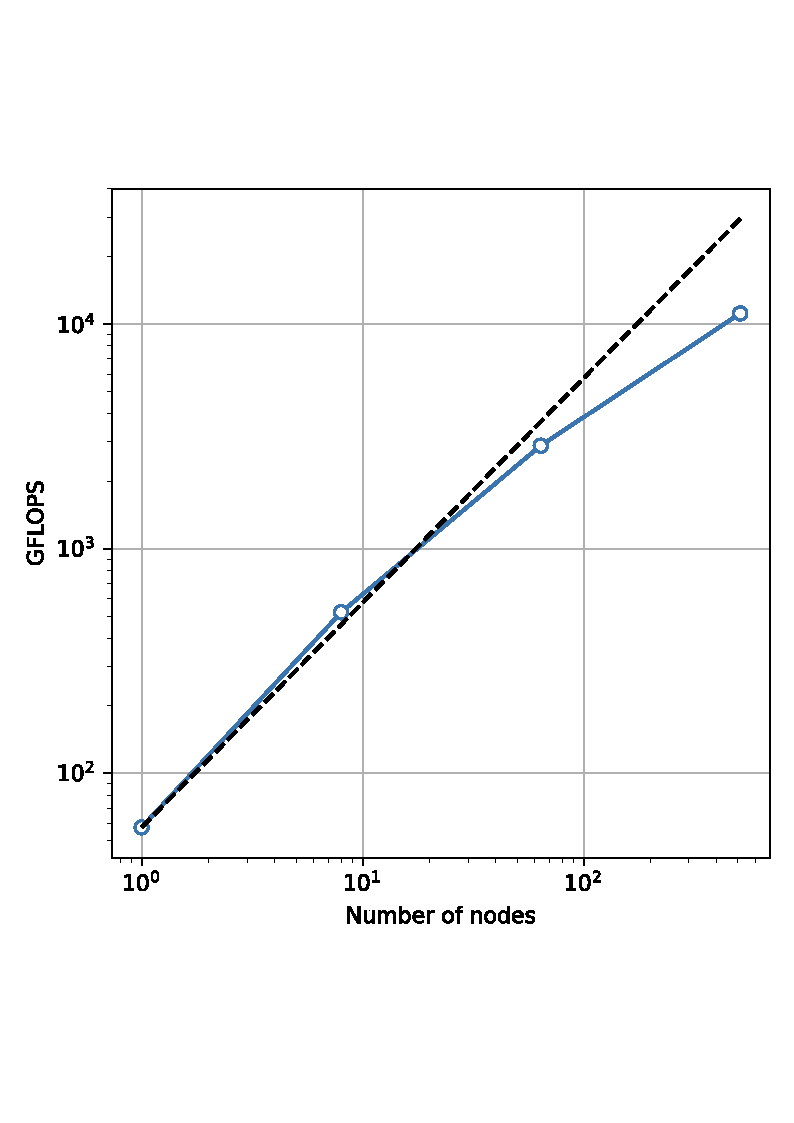}
	\caption{Weak scaling results of FrontFlow/red-HPC.}
	\label{fig:weak_scaling_ffr}
\end{figure}

Table \ref{table:weak_scaling_ffr} and Figure \ref{fig:weak_scaling_ffr} show the performance example of a CFD code {\bf FrontFlow/red-HPC} we have also developed for and optimized for Fugaku; it utilizes many of the modern techniques such as node-centered unstructured grid using the finite volume method tuned by data access reordering, Ellpack-Itpack generalized diagonal data storing and loop blocking for matrix-vector multiplication to improve effective performance etc. Although heavily optimized for Fugaku/A64FX, on the benchmark target model is NACA65-410 airfoil with the number of elements being 13,862,400. we achieve only 98.93 2.92\% of peak FLOP performance of 3 Teraflops, and 267.58 GB/sec.of memory bandwidth corresponding to 26.13\% of its peak of 1 Terabytes/s. Also, scaling falls off at fairly low number of nodes even with weak scaling, where we only achieve 37.94\% scaling at 512 nodes, primarily due to load imbalance.

\begin{table}[h]
\small\sf\centering
\caption{Weak scaling results of FrontFlow/red-HPC.\label{table:weak_scaling_ffr}}
\scalebox{0.95}{
\begin{tabular}{rrrrr}
\toprule
Nodes&Elements&Time&GFLOPS&Scaling\\
\midrule
1&1.73M&16.35 s&57 (1.70\%)&100.00\%\\
8&13M&13.45 s&521 (1.92\%)&113.35\%\\
64&110M&19.90 s&2,877 (1.33\%)&78.24\%\\
512&887M&45.20 s&11,164 (0.64\%)&37.94\%\\
\bottomrule
\end{tabular}}\\[10pt]
\end{table}

The most serious showstopper in applying such unstructured CFD methods is not only in the compute itself, but enormous time required in preprocessing to create high-resolution, high-quality geometries required. Typically in industry situations, this is conducted by experts who have years of experience in using the software tools required. By all means over the past 20 years, commercial CFD software packages have made remarkable progress in the preprocessing technology, such as easy-to-use grid generation software or emergence of surface wrapping technology. However, it still requires an expert several days to several weeks for grid preparation in a realistic setting as we discuss in Section~\ref{rapidmesh}. Although this may be fine for development of normal industry products such as automobiles, where the development time is in year and typically only minor alterations are made from the base geometry, not from scratch, in the case of COVID-19 droplet simulation this is totally insufficient, as completely different geometries representing the digital twins of various social situations must be generated in minutes, as 100s of simulations of different geometrical scenarios must be conducted to meet the demands from the government and the industry to curtail the pandemic.

\section{Innovations in our Approach in Modeling and Mitigating COVID-10 Droplet/Aerosol Transmissions}

The major innovations realized in in our approach in modeling and mitigating COVID-10 droplet/aerosol transmissions are exactly to overcome the critical shortcomings to realize physically accurate and extreme high-resolution droplet/aerosol simulation with very rapid turn around time, not just the simulation itself but the ability to generate the CFD models from CAD and scanned data very quickly, and extremely scalable on a modern exascale machines such as Fugaku, combined with the epidemiological infection models for accurate and on-time end-to-end assessment of infection risks. As far as we know no methodologies or software system existed to satisfy all of the requirements, but fortunately, transpositional adaptation of CUBE, developed for industrial CFD with similar requirements while being based on explicit methods for structured grids, for efficiency in both compute and mesh generation, but being able to handle difficult geometries such as both sides of window panes and bonnets of automobiles, would be able to satisfy the goals with its innovative features; and combined with epidemiological models and experiments, we are essentially transforming the ability of epidemiology to model COVID-19 and other types of future airborne diseases, whereas they have been restricted to experimentation and very simple mathematical modeling.

\subsection{Physical Modeling and Numerical Description of Aerosols and Droplets}

\subsubsection{Governing Equations for Flow}
The equations of motion in compact vector notation can be expressed as
\begin{equation}
	\frac{\partial \mathbf{U}}{\partial t}+\nabla \cdot \mathbf{F}=\mathbf{S}
	\label{eqn-geq}
\end{equation}
where the vectors $ \mathbf{U} $ and $ F $, given below, represent the primitive variables, and the convective and the diffusive terms, respectively. 

\begin{equation}
	\mathrm{\mathbf{U}}=\left(\begin{array}{c}
		\rho \\
		\rho u_{1} \\
		\rho u_{2} \\
		\rho u_{3} \\
		\rho e \\
		\rho Y_{k}
	\end{array}\right), F_{i}=\left(\begin{array}{c}
		\rho u_{i} \\
		\rho u_{i} u_{1}+P \delta_{i 1}-\mu A_{i 1} \\
		\rho u_{i} u_{2}+P \delta_{i 2}-\mu A_{i 2} \\
		\rho u_{i} u_{3}+P \delta_{i 3}-\mu A_{i 3} \\
		\rho(\rho e+P) u_{i}-\mu A_{i j} u_{j}+q_{i} \\
		\rho u_{i} Y_{k}-\rho \hat{u}_{i}^{k} Y_{k}
	\end{array}\right)
\label{eqn-UF}
\end{equation}

Here, the gas density is represented by $ \rho $ and the viscosity by $ \mu; u_{i}, e $ and $ P $ represent the velocity, the total specific energy, and the pressure, respectively. Species mass fraction and diffusion velocities of the $ k^{th} $ species are given by $ Y_k $ and $ \hat{u}_{i}^{k}$, respectively. The total specific energy is defined as $ e=P/(\gamma-1)+1/2 u_i u_i $ and $ q $ is the heat flux vector. The stress tensor is given by  $A_{i j}=\partial u_{j} / \partial x_{i}+\partial u_{j} / \partial x_{i}-2 / 3(\nabla \cdot u) \delta_{i j}$.  The components of the source term are the following $\mathbf{S}=\left(\rho-\rho_{0}\right)\left[0, g_{1}, g_{2}, g_{3}, g_{i} u_{i}, S_{\rho Y_{k}} /\left(\rho-\rho_{0}\right)\right]$, where $ \rho_0  $ is the far field ambient density, $ \rho $ is the local density, g is the acceleration due to gravity (eg. $ g=(0,0,-9.81)m/s^2 $) and $ S_(\rho Y_k ) $ is the source term due to droplet evaporation. The viscosity and thermal conductivity are treated as a function of temperature based on Sutherland’s Law. Lastly, the density and pressure are constrained together by the state equation $ P=\rho RT $, in which $ R $ is the gas constant and $ T $ is the temperature. 

For time marching, the dual time stepping scheme of  \cite{weiss1995} is adopted and the governing equations are reformulated as 

\begin{equation}
	\Gamma \frac{\partial \mathbf{U}_{p}}{\partial t}+\frac{\partial \mathbf{U}}{\partial t}+\nabla \cdot \mathbf{F}=\mathbf{S}
	\label{eqn-implicit-geq}
\end{equation}
where $ \mathbf{U}_p $ has the primitive form $ [P,u_1,u_2,u_3,T] $ and $ \Gamma $ is the preconditioning matrix \cite{weiss1995}. Further details of the spatio-temporal discretization can be found in the work of \cite{li2015}. The implicit duel time stepping scheme is applied separately to the fluid flow equations and the species transport equations and not applied to the combined system.  

\subsubsection{Discretization of Governing Equations}

The discretized form of Eq.~\ref{eqn-implicit-geq} is
\begin{equation}
    \begin{aligned}
        &\Gamma \frac{\bar{U}_{p}^{k+1}-\bar{U}_{p}^{k}}{\Delta \tau}+\frac{3 \bar{U}^{k+1}-4 \bar{U}^{n}+\bar{U}^{n-1}}{2 \Delta t} \\
        &+\frac{1}{\Delta \xi}\left(\bar{F}_{1+\frac{1}{2}, j, k}^{k+1}-\bar{F}_{1-\frac{1}{2}, j, k}^{k+1}\right)\\
        &+\frac{1}{\Delta \eta}\left(\bar{F}_{i, j+\frac{1}{2}, k}^{k+1}-\bar{F}_{i, j-\frac{1}{2}, k}^{k+1}\right)\\
        &+\frac{1}{\Delta \zeta}\left(\bar{F}_{3, j, k+\frac{1}{2}}^{k+1}-\bar{F}_{3, j, k-\frac{1}{2}}^{k+1}\right)=\bar{S}
    \end{aligned}
\label{eqn-discrete}
\end{equation}
The superscripts $ k $ and $ n $ indicate the iteration numbers of artificial time and the proceeding step of real time, respectively.

And then, Eq.~\ref{eqn-discrete} can be rearranged as

\begin{equation}
    \begin{aligned}
&   \left[\frac{I}{\Delta \tau}+\Gamma^{-1} M \frac{3}{2 \Delta t}+\Gamma^{-1}\left(\delta_{\xi} A_{p}^{k}+\delta_{\eta} B_{p}^{k}+\delta_{\zeta} C_{p}^{k}\right)\right]\\
&  \Delta U_{p}=\Gamma^{-1} R^{k}
    \label{eqn-discrete-time}
   \end{aligned}
\end{equation}
where  $ M = \partial U/\partial U_p, \delta_{x_i} $, the central-difference operator, $ A_p = \partial F_{1}^{k}\partial U_p $ the flux Jacobian, and  $R^{k}=-\left(3 U^{k}-4 U^{n}+U^{n-1}\right) /(2 \Delta t)-\left(\delta_{x_{1}} \bar{F}_{1}^{k}+\delta_{x_{2}} \bar{F}_{2}^{k}+\delta_{x_{3}} \bar{F}_{3}^{k}\right)$.  Eq.~\ref{eqn-implicit-geq} is solved using the Lower-Upper Symmetric-Gauss-Seidel (LUSGS) implicit method.

In the calculation of   on the riht-hand side of Eq.~\ref{eqn-discrete-time}, the terms involving   in Eq.\ref{eqn-implicit-geq} can be divided into an inviscid term $F_{inviscid}$  and a viscous term $ F_{viscous}$. The Roe scheme is employed in discretizing $ F_{inviscid}$,

\begin{equation}
    F_{\text {inviscid }, i+1 / 2}=\frac{1}{2}\left[F_{R}(U)+F_{L}(U)\right]+F_{d}
    \label{eqn-Fsplit}
\end{equation}
where $ F_d $ is the Roe dissipation term, which is composed of jumps of properties of work ﬂuids. At extremely low Mach numbers, the dissipation term is modified as the Preconditioning-Roe \cite{weiss1995} to resolve Eq.~\ref{eqn-Fsplit}. For the reconstruction of $ F_R $ and $ F_L $, the fifth-order MUSCL scheme (the monotonic upstream-centred scheme for conservation laws) \cite{li2015} is adopted. Aside from the inviscid term, the derivative terms in $ A_{ij} $ in the viscous term of Eq.~\ref{eqn-discrete-time} are calculated using the second-order central difference.

\subsubsection{Numerical Modeling of Droplets/Aerosols}

For modeling sputum droplets, we adopt the single droplet model which is widely used in the literature. The droplets are treated as discrete Lagrangian particles. The transport of the droplets is modeled by the following equation. 

\vspace{-10pt}
\begin{equation}
\frac{d \mathbf{u}_{d}}{d t} =\frac{3 C_{D}}{4 d_{d}} \frac{\rho}{\rho_{d}}\left(\mathbf{u}-\mathbf{u}_{d}\right)\left|\mathbf{u}-\mathbf{u}_{d}\right|+\mathbf{g} ,
\label{eqn-dropletU}
\end{equation}

\vspace{-10pt}
In the above equations, $ u_d $ is the droplet velocity, $ \rho_d $ is the droplet density, $  C_D $ is the drag coefficient expressed a function of the droplet Reynolds number $ Re_d $ and $ d_d $ is the droplet diameter. The evaporation of the droplets is modeled by tracking the temperature and mass of the droplets.  

\vspace{-10pt}
\begin{equation}
\begin{aligned}
\frac{d T_{d}}{d t} &=\frac{N u}{3 P r} \frac{c_{p}}{c_{l}} \frac{f_{1}}{\tau_{d}}\left(T-T_{d}\right) + \frac{1}{m_{d}}\left(\frac{d m_{d}}{d t}\right) \frac{L_{V}}{c_{p, d}} \\
\frac{d m_{d}}{d t} &=-\frac{m_{d}}{\tau_{d}}\left(\frac{S h}{3 S c}\right) \ln \left(1+B_{M}\right)
\end{aligned}
\end{equation}
Here, $ m_d $, and $ T_d $ are the mass and temperature, respectively. $ T $ is the temperature of the ambient air, $ L_V $ the latent heat of evaporation at the droplet temperature. $ c_p $ and $ c_l $ are the specific heat at a constant pressure of the ambient air and the specific heat capacity of the liquid droplet, and $ \tau_d $ is the response time of the droplet. $ f_1 $ is a correction to the heat transfer due to evaporation of the droplet. $ Nu \& Pr $ are the Nusselt and Prandtl numbers, respectively. And, $ Sh, \;Sc \; \& \; Bm $ are the Sherwood number, Schmidt number, and the mass transfer number, respectively. In order to model the effect of nucleation of sputum droplets due to the presence of viral particles and salts, the droplet evaporation calculation is stopped when the droplet radius reaches $ 0.5\;\mu$m. Further details of the various terms involved in the droplet model can be found in the work of \cite{bale2021}.

\subsection{Rapid Creation of Digital Twins of Societal Situations for Droplet/Aerosol Assessment}

In addition to simulating the droplets/aerosols numerically, there must be methods to quickly incorporate geometry and other physical models of societal situations as digital twins for aerosol/droplet simulations. This requires conversion of design data from CAD, or scanned data from the physical environment, to converted to high resolution surface geometry data for CFD.

For the assessment of public transportation such as buses, taxis, commuter trains, and aircrafts, etc., as well as common public places such as offices and concert halls, digital CAD data are present for CAE applications, but the overall design pipeline is intended to use traditional unstructured meshes for accuracy. This has a significant effect on the overall turn-around time, as geometry creation is extremely costly, especially with the additions of other components in the digital twin, e.g., people, furniture, engines, and various pipes \& ducts in cars, not just body surfaces, as well as various anti-COVID measurements such as shields, curtains as well as people wearing protective gears of different varieties, etc. Here, our new method allows geometry generation in minutes instead of weeks---this will be discussed in \ref{rapidmesh} and \ref{BCM}.

Although not typical, we also devised methodologies for the assessment of a specific, actual public rooms such as hospital wards, classrooms, live music clubs, etc., where such CAD data do not exist. In such cases, we have utilized point (cloud) digitization technique in which we place on site 3D laser scanning system and create the point group (cloud) data of the target room shape, from which we create the surface CAD data for the simulation, and combine it with artifacts such as ducts for which we do have CAD data. 3D laser scanning takes less than one hour, while a few days are required to convert the point cloud data to surface CAD data. However, once the geometry is generated, it can be combined with other geometries in various social settings such as people and table placements, and mesh generation for such variances can be done in a manner of minutes.

\begin{figure*}
\centering
\includegraphics[keepaspectratio,width=\textwidth]{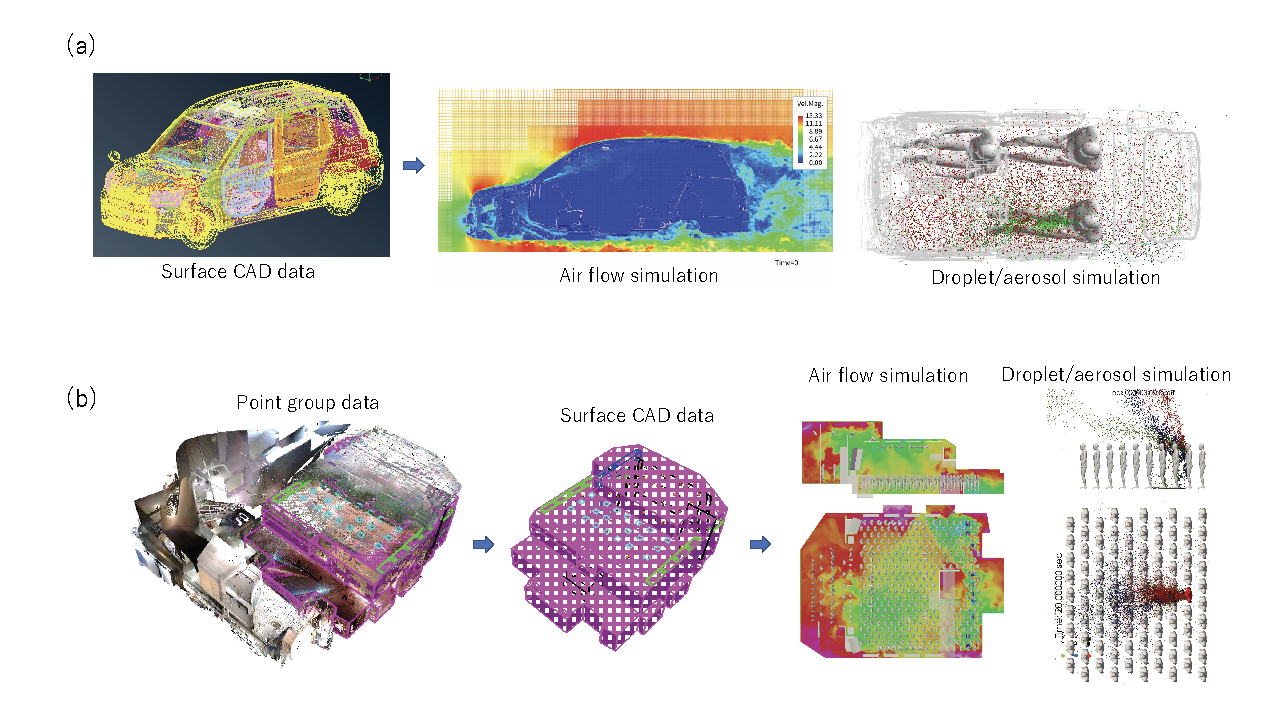}
\caption{Digital transformation of the viral infection risk assessment: (a)In the case of CAD data of the target is existing, (b)In the case digital data of the target geometry does not exist.\label{fig:modeling}}
\end{figure*}

\subsection{Rapid Mesh Generation Based on the Immersed Boundary Method}
\label{rapidmesh}

One of the core innovations we have realized to achieve significant reduction in turnaround time in developing large-scale parallel computing method is to combine the hierarchical Cartesian grid and the immersed boundary method (IBM), which does not require preprocessing work \cite{onishi2021}.

\begin{figure}[H]
  \centering
  \begin{minipage}{0.4\hsize}
    \includegraphics[keepaspectratio,width=\textwidth]{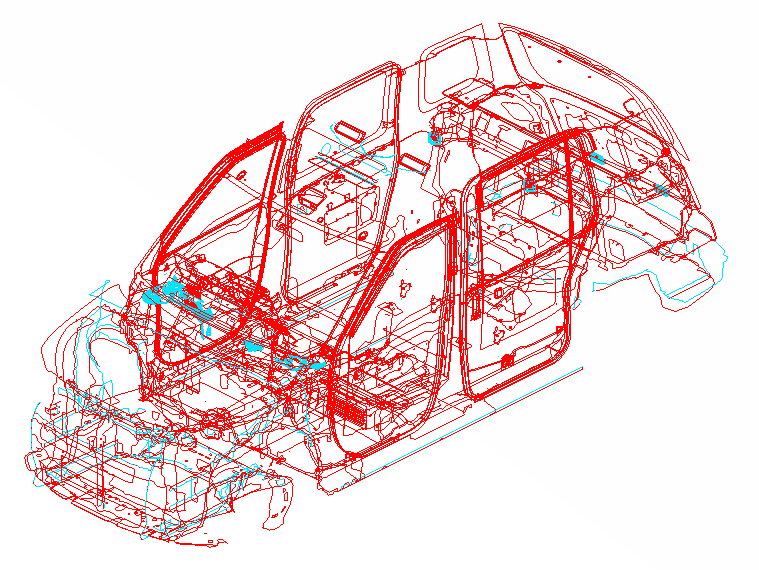}
    \subcaption{}
    \label{fig:geom_error}
  \end{minipage}
  \begin{minipage}{0.5\hsize}
    \includegraphics[keepaspectratio,width=\textwidth]{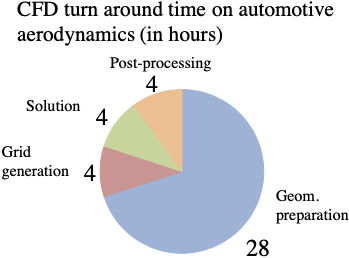}
    \subcaption{}
    \label{fig:CFD_turnaround}
  \end{minipage}
  \begin{minipage}{0.5\hsize}
    \includegraphics[keepaspectratio,width=\textwidth]{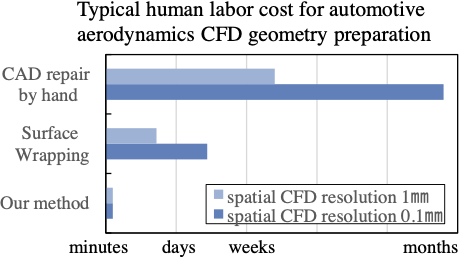}
    \subcaption{}
    \label{fig:CFD_geom_pre}
  \end{minipage}
  \caption{(a) Problematic area of geometric data. Red lines indicate gaps where two faces are not connected; light blue lines indicate edges that are over-connected. (b) Example of turnaround in CFD on vehicle aerodynamics analysis by \cite{furukawa2004application} and (c) geometry preparation time.}
  \label{fig:dirty_CAD}
\end{figure}

The difficulty in grid generation, especially for unstructured grids, is mainly due to the inconsistency contained in the geometry data. For example, gaps, overlaps, and surface defects that occur during CAD data conversion hinder the creation of closed spaces for grid generation. Further,  superposition of numerous thin plate structures, the surface existing inside each part and the surface without thickness, etc,  hiders the inside/outside judgment of a fluid region. Moreover, the small shapes and features below grid resolution hinders grid generation process itself. Such data is called ''dirty'' CAD data; Figure \ref{fig:dirty_CAD}\subref{fig:geom_error} shows the geometry model which is used in ventilation analysis for a taxi. The red and light blue lines indicate the edges that the user needs to repair manually after being processed through a conversion tool, including separated edges that should be connected, and over-connected edges that should be separated. There are about 400,000 such error points. Note further that, even with such corrections the geometry is not ‘watertight’ and it is required to delete unnecessary shapes and simplify them before creating the grids.

To avoid this problem, the idea is to handle the geometry separately from the calculation grid. In this regard, IBM was developed by \cite{Peskin1972} et al. for moving boundary problems, and refers to a method of calculating the existence of a shape by replacing with the external forcing term in the governing equation instead of representation with the grid shape itself. The method has been successfully applied in fields such as a heart valve pulsation or linear object locomotion and cardiovascular vessels, but the technique has also been known to embody difficulties in severe time-step restrictions. Later on, \cite{Mittal2007} et al. extended the method and evolved it to be applicable to sharp shapes. Based on such past work, we designed a method to solve the problem of dirty CAD by combining the essence of these two work; Peskin et al. formulated the external forcing by dispersion operation, while Mittal et al. formulated it as a boundary condition for velocity field inversion. Both are expressed in general form as Eq. \eqref{eq:IBM_forcing},

\begin{align}
  \label{eq:IBM_forcing}
  \mathcal{F}(\bm{x},t) &= \sum_{k} \bm{f}_{k}^{n+1/2} \delta (\mid \bm{x} - \bm{X}(s,t) \mid) & \nonumber \\
  & \approx \sum_{k} \bm{f}_{k}^{n+1/2} \mathcal{D} (\mid \bm{x} - \bm{X}(s,t) \mid), \forall \bm{X}(s,t) \in \Gamma, &
\end{align}

\begin{align}
  \label{eq:IBM_distribution}
  & \mathcal{D} (\mid x - X \mid) =
  \begin{cases}
\displaystyle  \int_{\phi} G(\mid x - X \mid) \delta (\mid x - X \mid), &\\
 \qquad for \; x \in \Omega_{lagrangian}, &\\
\displaystyle  \zeta \left[ V \right]^{-1} [ q ] + \eta, &\\
 \qquad for \; x \in \Omega_{fluid}, &\\
\displaystyle  \zeta \sum_{i=1}^{m} \alpha_{i} \, q_{i} + \eta, &\\
 \qquad for \; x \in \Omega_{fluid+ghost}, & 
  \end{cases} &
\end{align}
where $\mathcal{F}$ represents the discrete force and $\mathcal{D}$ represents the distribution function of a vector function $\bm{X}(s,t)$ that gives the location of points of boundary $\Gamma$ for position $s$ and time $t$. Here, there are several approaches to implement $\mathcal{D}$---one is to have a dummy cell in the grid including a wall surface (i.e., $\Omega_{fluid+ghost}$, axis projected interpolation); another is to treating the wall surface as a representative point cloud discretized by a background grid ($\Omega_{\it lagrangian}$, Gaussian operator interpolation), and their combination thereof ($\Omega_{fluid}$, interpolation using Vandermonde matrix $V$). The commonality of these approaches is to avoid the difficulty of determining the inside and outside of fluids due to the thin surfaces by locally dividing the fluid/solid region. This simple improvement yields critical benefits that will not cause errors in any types of geometry data, and is applicable to Cartesian grids. We have applied the variations of these methods to various industrial problems \cite{Jansson2019,Onishi2018,Li2016,bale2020a,Cao2019,Lu2021}. For COVID-19 digital twins of societal situations, the optimal method was selected for one making it possible to automatically and robustly simulate complex dirty geometry, reducing the time required for preprocessing to about 10 minutes or less.

\subsection{Hierarchically Structured Grid System by Building Cube Method}\label{BCM}

Although our evolved IBM method allows for use of regular, Cartesian grids for good locality and scalability, for industry problems we will suffer immense load imbalance and significant waste in terms of not only CPU but memory as well, especially for very fine geometries. However, simple hierarchical methods such as octree results in too many decompositions and inefficiency. Here, we employ a hierarchical Cartesian grid method called the Building Cube Method or BCM, proposed by \cite{Nakahashi2003}, which takes into account the architectural properties of modern massively parallel supercomputer systems. 

The basic data structure is similar to that used in adaptive mesh refinement. BCM assumes the grid to be constructed as hierarchical equidistant Cartesian coordinates with two fixed levels of subdivided spatial domains. The computational space is first divided into ''Cubes'', which are split in an octree fashion as they approach the walls of the geometry. Figure \ref{fig:BCM_grid} shows an example of a grid. At the leaf, a Cube is divided into ''Cells'', typically $16 \times 16 \times 16$ are assigned in 3D, irrespective of the spatial resolutions. Since each Cube is composed of the same number of cells, it is easy to evenly distribute the computational load to each process and achieve high parallelization and load balancing. 

The interface of each Cube is discontinuous between adjacent Cubes, so it is necessary to exchange data between them. Each Cube has a buffer area (halo Cell) at the interface, and data is exchanged between these buffers. The exchange is not a simple transfer of halo data, but rather, the interface information is computed at the same time. One option is a first-order averaging interpolation from a small Cube to a large Cube, and the other is to perform 0th order interpolation from a large Cube to a small Cube and between the same size Cubes (in this case it becomes a copy operation). Data is exchanged diagonally as well as along its axis when required. As is with other tree-based techniques, although the order of data access is made sequential since the data buffer area becomes discontinuous for performance degradation. Thus, compute/communication overlap is employed to conceal the overhead. Execution performance is shown in the section \ref{performance_BCM}.

The modeling of droplet dispersion is also based on discretization in Cube units. The Lagrangian marker particles are grouped into sets matching the underlying Cube data-structure. The insertion and removal of Lagrangian particles, as they are transported from one Cube to another, is carried out using the particle’s global index-based hash table. This reduces the number of searches between the Lagrangian particles and the Euler grids for high parallel efficiency. Particles that reach the Cube interface are exchanged via the buffer region and transferred to the next Cube.

\subsection{Mixed Lagrangian/Eulerian Formulation}

Numerical formulation employed for modeling droplet dispersion in this work consists of a combination of a Lagrangian and an Eulerian frame of reference. The discretized equations of mass, momentum and energy conservation reside on the Eulerian meshed. The equations for conservation of evaporation phase of liquid droplets and species of the gas phase are also solved on the Eulerian mesh. The liquid droplet dynamics equations are solved on the Lagrangian frame and the immersed boundary geometry is represented as Lagrangian entity to model moving geometries. 

\begin{figure}[H]
  \centering
  \begin{minipage}{0.2\hsize}
    \includegraphics[keepaspectratio,width=\textwidth]{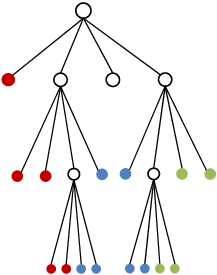}
    \subcaption{}
  \end{minipage}
  \begin{minipage}{0.35\hsize}
    \includegraphics[keepaspectratio,width=\textwidth]{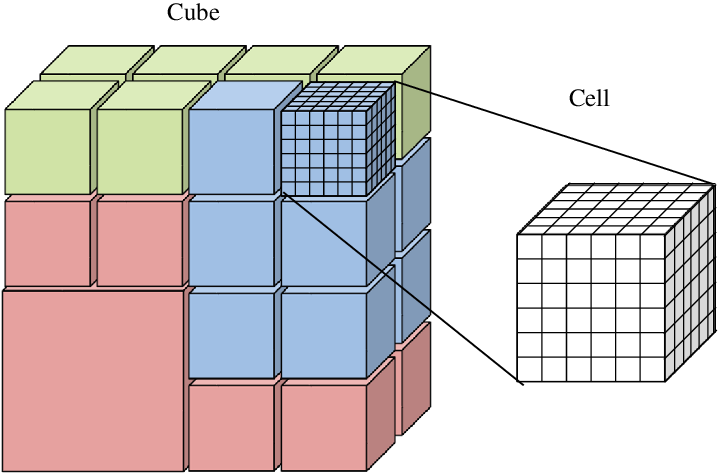}
    \subcaption{}
    \label{fig:BCM_grid}
  \end{minipage}
  \begin{minipage}{0.4\hsize}
    \includegraphics[keepaspectratio,width=\textwidth]{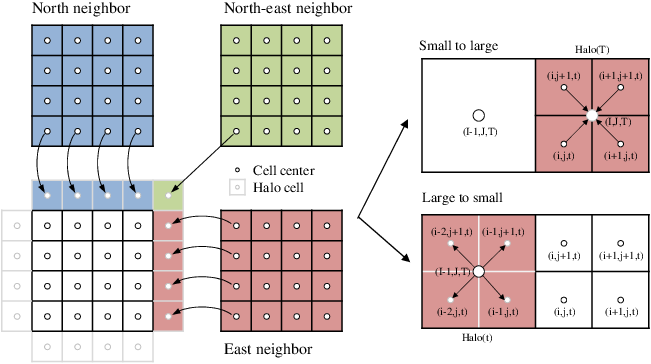}
    \subcaption{}
    \label{fig:BCM_halo}
  \end{minipage}
  \begin{minipage}{0.3\hsize}
    \includegraphics[keepaspectratio,width=\textwidth]{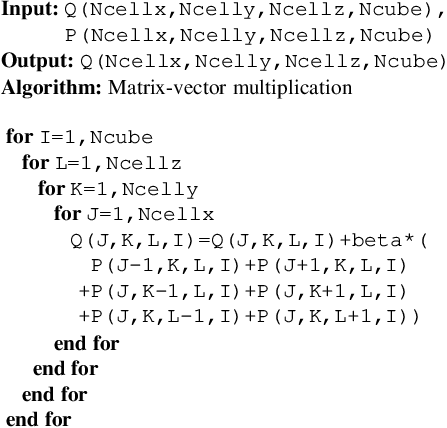}
    \subcaption{}
    \label{fig:BCM_loop}
  \end{minipage}
  \begin{minipage}{0.2\hsize}
    \includegraphics[keepaspectratio,width=\textwidth]{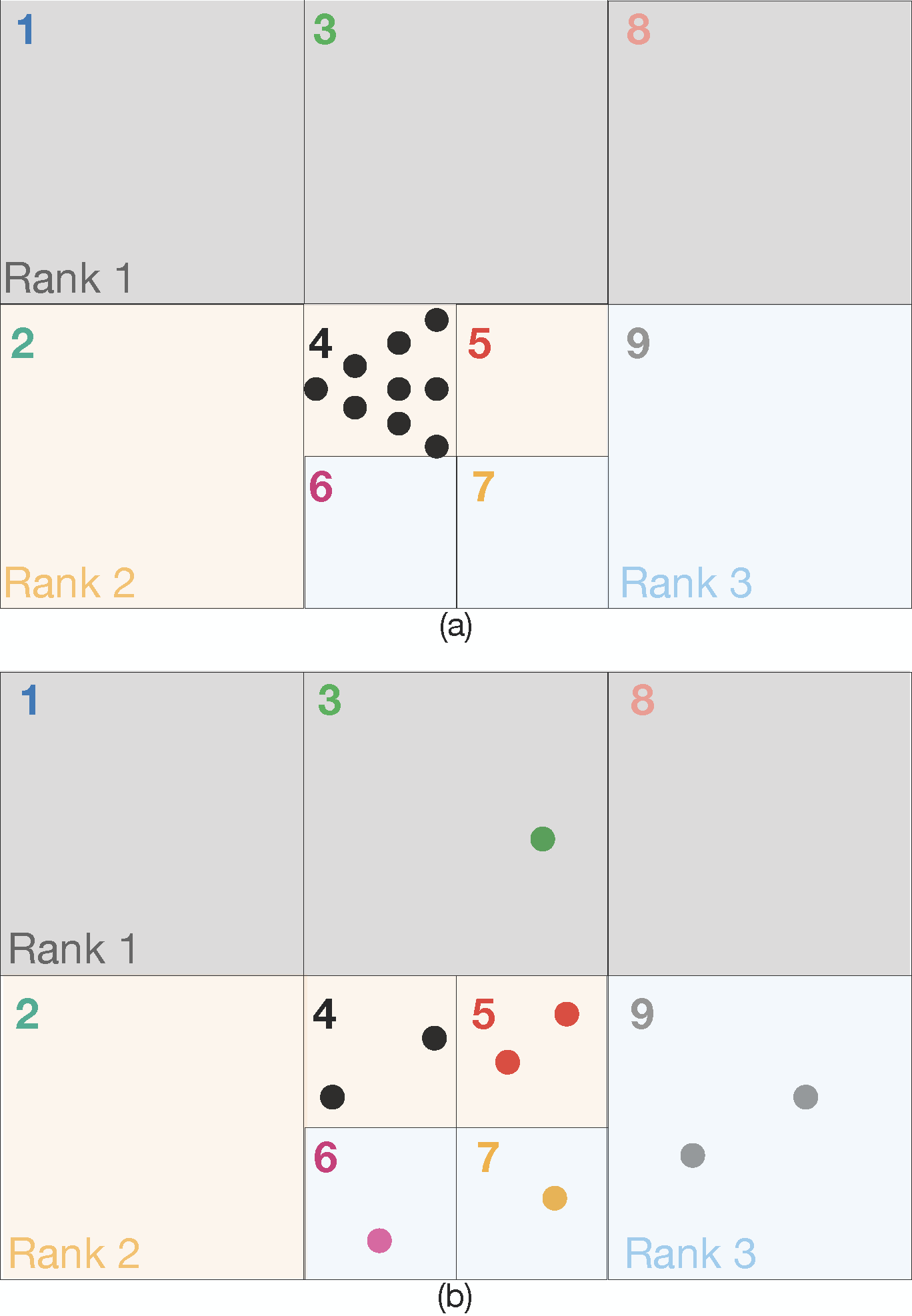}
    \subcaption{}
    \label{fig:BCM_lag}
  \end{minipage}
  \caption{(a) Octree like data structure of generating Cubes, (b) exchange of information between neighboring Cubes on halo cells and (c) the typical loop structure for BCM grids. (d) Allocation of Lagrangian particles to respective Cube as they move from one to another.}
  \label{fig:BCM}
\end{figure}

The indexing of BCM mesh’s Cube blocks is arbitrary. Therefore, identifying the Cube in which a given Lagrangian marker particle is situated would involve search operations, which is preferred to be avoided when possible, for efficiency. Furthermore, depending on the application, the total number of droplets and consequently the number of Lagrangian marker particles can be very large, as high as $ O(10^7) $. Consequently, care must be taken in developing the Lagrangian data structure to optimize computational performance. For this, we employ the same strategy of discretization as is used in the BCM mesh. During initialization, the Lagrangian marker particles are grouped into sets wherein the sets are defined to match the underlying Cube data-structure of the BCM mesh. Each set of the Lagrangian data structure is defined such that only those Lagrangian marker particles that are within the corresponding Cube of the BCM are assigned to the set. The insertion and removal of Lagrangian particles, as they are transported from one Cube to another (Fig.~\ref{fig:BCM_lag}), is efficiently carried out using the particle’s global index-based hash table. With regards to the domain decomposition for parallel computing, a spatial decomposition strategy is employed for the combined Lagrangian-Eulerian system.

\subsection{End-to-End Modeling of COVID-19 Transmissions}

 \begin{figure}
    \centering
    \includegraphics[width=0.5\textwidth]{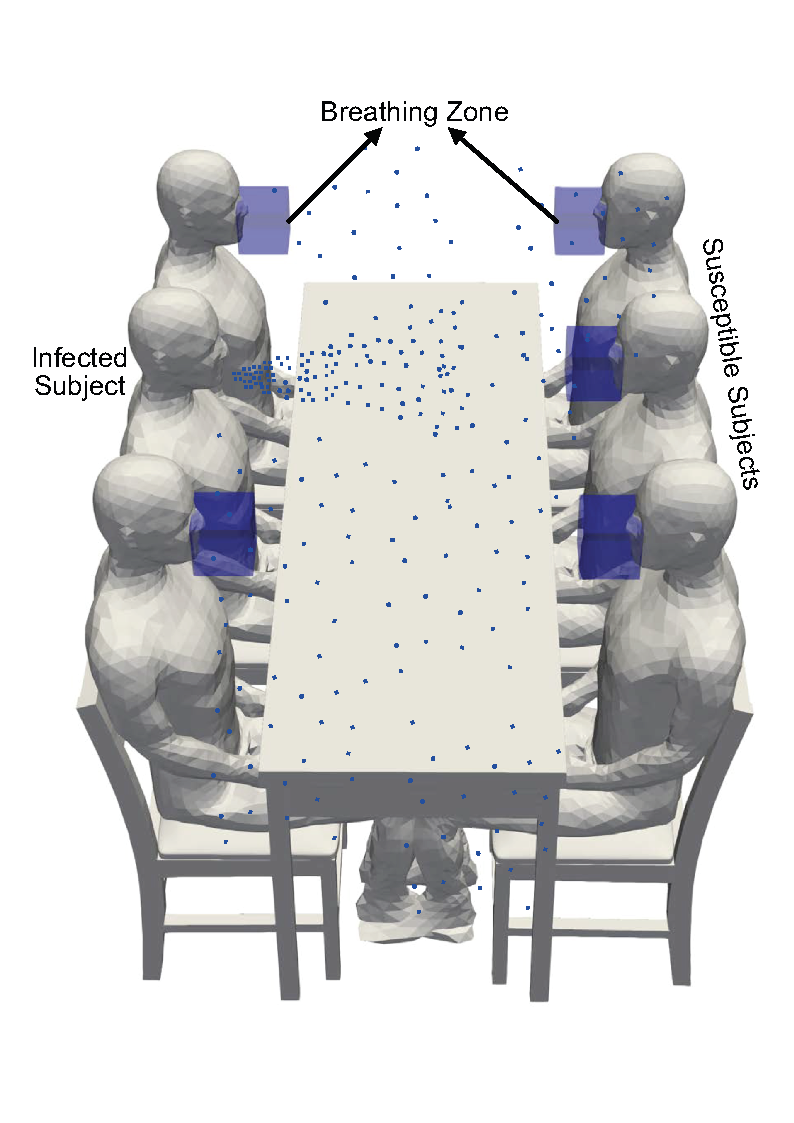}
    \caption{A schematic of the methodology employed for estimating virions within the breathing zone of an infected person. \label{fig:breathing-zone}}
\end{figure}

The current and the first-version of our end-to-end transmission model deals with the amount of virus exhalation and inhalation by the subjects. Although simple, this is consistent with epidemiological experimental practices.

The number of viral particles or virions that are likely to be inhaled by a susceptible person who has been exposed to an infected environment for a duration of T can be evaluated as follows.
\begin{equation}
    N = \frac{B}{v_B}\int_{0}^{T} n(t)dt,
    \label{eqn:N_count_1}
\end{equation}
where, $ B $ is the breathing rate, $ n(t) $ is the local count of virion within the breathing zone of the susceptible subject, and $ v_B $ is the volume of the breathing zone. As the focus of our work is transient evolution of droplet dispersion, the number of virions at any given location can vary with time, therefore it is expressed as a function of time $ t $. The breathing zone is a region within the vicinity of the mouth and nose of a susceptible person from which air is likely to be inhaled. The size and dimensions of the breathing zone can be determined through direct numerical simulation of the inhalation process. In this work however, we choose a rectangular box located in front of and below the nose.

The virion count depends on the number of droplets and aerosols that enter the breathing zone. It may be expressed as a product of the ejection volume of a droplet $ v_{d}^{0} $ and the viral load of SARS-CoV-2 $ \rho_v $. Therefore, the final form of $ N $  can be written as
\begin{equation}
    N =  \frac{B\rho_v}{v_B}\int_{0}^{T} v_{d}^{0}(t)dt.
\end{equation}
For quasi-steady state situations the integral term in the above equation can be replaced with
\begin{equation}
    N =  \frac{B\rho_v}{v_B}\bar{v}_{d}^{0}T,
\end{equation}
where $ \bar{v}_{d}^{0} $ is the quasi-steady state rate of droplet ejection volume entering the breathing zone. Further details of the model can be found in the work of \cite{bale21arxiv}

We are currently working to refine this model to deal with aerosols that are directly generated inside the lung, exhaled, and then inhaled into the lung and actually infection the cells. The results from such work will be reported in our future publications.

\section{Performance Measurement}

The supercomputer Fugaku was mainly used for performance evaluation\cite{Yoshida2018-ag, R-CCS2021}. At the time of this writing, it is the fastest supercomputer in the world according to the TOP500 list\cite{top500_2021} as well as other benchmark lists, HPCG, HPL-AI, and the Graph500 for three consecutive editions, in June and November 2020, as well as June 2021 \cite{RIKEN2021}.
Each node of Fugaku is a general-purpose CPU, the A64FX™ CPU, developed by Riken and Fujitsu. The A64FX™ CPU supports the Armv8.2-A SVE which is a new extension supporting 512-bit vector instructions. 
Although being Arm-compliant, the microarchitecure of each A64FX™ core is original, optimized for HPC with SVE and high bandwidth, running at 2.0 GHz in normal mode and 2.2 GHz in boost mode. 13 compute cores are clustered into a core memory group (CMGs), with one core being an OS-dedicated assistant-core, while 12 are compute cores, all of them mutually interconnected as well as to a 8MB L2 which in turn is connected to an HBM2 module providing 256GB/s of bandwidth. There are 4 CMGs in each CPU chip, interconnected with each other via an on-chip ring-bus, which also connects to a on-chip scalable system interconnect TOFUD, with embedded switches. The 48 compute cores are cache-coherent, constituting a on-chip NUMA system, with 32MB of shared L2 chache, and 32 GB of HBM2 with 1024GB/s of bandwidth, several times faster than standard server memory system.

The peak arithmetic performance of the A64FX™ in normal mode is 3.072 TFLOPS for double-precision, 6.144 TFLOPS for single-precision, and 12.288 TFLOPS for half-precision. In boost mode, these increase to 3.3792 TFLOPS, 6.7584 TFLOPS, and 13.5168 TFLOPS, respectively. Efficiency is very high, with DGEMM achieving over 90\% of peak, and the Stream Triad performance is over 830 GB/sec or 81\% of peak\cite{Fujitsu2019}. The TofuD interconnects constitutes a six-dimensional torus, with 6.8 GB/sec per link, and with six Tofu network interfaces (TNIs) eminating from the chip out of its embedded intra-chip switch, injection bandwidth is 40.8 GB/sec per node. In total,
Fugaku has 158,976 nodes and 7,630,848 cores. Theoretical system peak performance is 488 PFLOPS in normal mode and 537 PFLOPS in boost mode \cite{R-CCS2021} and 163 PByte/s of memory bandwidth; we use the former for the measurement against peak as CUBE is largely memory bound and does not benefit from higher clock frequency.


\subsection{Single-node performance}\label{singlenode_performance}


As a single node performance benchmarks we employ the cavity flow in a region that extends along the y-axis direction. The domain is decomposed in one-dimension for ideal neighbor communication between each MPI process.
Table \ref{table:time_integ_loop_performance} shows the performance of the entire program (time integration loop) for A64FX; the performance of one node with 4 MPI processes, each assigned to a CMG, is 3.59 times that of 1 CMG (1 MPI process). Double-precision floating-point performance is 59.96GFLOPS for 1 CMG and 215.84GFLOPS for one node, which is 7.81\% and 7.03\% of the peak, respectively, relatively high values for industry-grade CFD performance, and much higher than 3.29\% in HPCG (16PFLOPS out of 488PFLOPS peak for normal mode) for Fugaku.

Table \ref{table:perf_comparison} is the performance comparison between Fugaku and Intel processors. The target problem is the cavity flow simulation the same above. As we observe, Fugaku/A64FX is substantially faster than Xeon and Xeon Phi platforms on a per CPU socket basis, being approximately 3x better in efficiency to peak as well as Performance/Watt basis, due to extreme high bandwidth and high power efficiency.

\begin{table}[h]
\small\sf\centering
\caption{Time integration loop performance.\label{table:time_integ_loop_performance}}
\scalebox{0.95}{
\begin{tabular}{rrrr}
\toprule
&Time&Calc. performance&Mem. BW\\
\midrule
1 CMG&38.40 s&59.96 GFLOPS&48.44\\
&&(7.81\%)&(18.92\%)\\
1 node&10.67 s&215.84 GFLOPS&171.90\\
&&(7.02\%)&(16.78\%)\\
\bottomrule
\end{tabular}}\\[10pt]
\end{table}

Table \ref{table:perf_comparison} is the performance comparison between Fugaku and Intel processors. The target problem is the cavity flow simulation the same above. As we observe, Fugaku/A64FX is substantially faster than Xeon and Xeon Phi platforms on a per CPU socket basis, being approximately 3x better in efficiency to peak as well as Performance/Watt basis, due to extreme high bandwidth and high power efficiency.

\begin{table}[h]
\small\sf\centering
\caption{Performance comparison between Fugaku and Intel Xeon systems (Fujitsu's commodity server, Oakforest-PACS\cite{Oakforest-PACS}, and Oakbridge-CX\cite{Oakbridge-CX}).\label{table:perf_comparison}}
\scalebox{0.75}{
\begin{tabular}{p{3.0cm}p{2.7cm}p{2.7cm}p{2.7cm}p{2.7cm}}
\toprule
&Fugaku&Intel server&Oakforest-PACS&Oakbridge-CX\\
\midrule
\multicolumn{5}{l}{\textbf{Specification}}\\
Node&Fugaku&Fujitsu PRIMERGY RX2540 M5&Fujitsu PRIMERGY CX1640 M1&Fujitsu PRIMERGY CX2550 M5\rule[0mm]{0mm}{3mm}\\
CPU&A64FX&Intel Xeon Gold 6240&Intel Xeon Phi 7250&Intel Xeon Platinum 8280\\
Clock&\multicolumn{1}{r}{2.0 GHz}&\multicolumn{1}{r}{2.6 GHz}&\multicolumn{1}{r}{1.4 GHz}&\multicolumn{1}{r}{2.7 GHz}\\
\# of CPUs&\multicolumn{1}{r}{1}&\multicolumn{1}{r}{2}&\multicolumn{1}{r}{1}&\multicolumn{1}{r}{2}\\
\# of cores&\multicolumn{1}{r}{48}&\multicolumn{1}{r}{36}&\multicolumn{1}{r}{68}&\multicolumn{1}{r}{56}\\
Peak&\multicolumn{1}{r}{3,072 GFLOPS}&\multicolumn{1}{r}{2,995 GFLOPS}&\multicolumn{1}{r}{3,046 GFLOPS}&\multicolumn{1}{r}{4,383 GFLOPS}\\
Peak&\multicolumn{1}{r}{1,024}&\multicolumn{1}{r}{281.6}&\multicolumn{1}{r}{490.0}&\multicolumn{1}{r}{281.6}\\
Memory bandwidth&\multicolumn{1}{r}{GB/s}&\multicolumn{1}{r}{GB/s}&\multicolumn{1}{r}{GB/s}&\multicolumn{1}{r}{GB/s}\\
\midrule
\multicolumn{5}{l}{\textbf{Execution setting}}\\
Processes (/node)&\multicolumn{1}{r}{4}&\multicolumn{1}{r}{2}&\multicolumn{1}{r}{2}&\multicolumn{1}{r}{1}\rule[0mm]{0mm}{3mm}\\
Threads (/node)&\multicolumn{1}{r}{48}&\multicolumn{1}{r}{36}&\multicolumn{1}{r}{48}&\multicolumn{1}{r}{48}\\
\midrule
\multicolumn{5}{l}{\textbf{Measured value}}\\
Execution time&\multicolumn{1}{r}{11.19 s}&\multicolumn{1}{r}{20.61 s}&\multicolumn{1}{r}{26.94 s}&\multicolumn{1}{r}{8.29 s}\rule[0mm]{0mm}{3mm}\\
Performance&\multicolumn{1}{r}{215.84 GFLOPS}&\multicolumn{1}{r}{66.01 GFLOPS}&\multicolumn{1}{r}{50.50 GFLOPS}&\multicolumn{1}{r}{164.13 GFLOPS}\\
(Peak ratio)&\multicolumn{1}{r}{(7.03\%)}&\multicolumn{1}{r}{(2.20\%)}&\multicolumn{1}{r}{(2.35\%)}&\multicolumn{1}{r}{(3.96\%)}\\
Memory bandwidth&\multicolumn{1}{r}{164.19 GB/s}&\multicolumn{1}{r}{49.63 GB/s}&\multicolumn{1}{r}{75.89 GB/s}&\multicolumn{1}{r}{85.82 GB/s}\\
(Peak ratio)&\multicolumn{1}{r}{(16.03\%)}&\multicolumn{1}{r}{(17.62\%)}&\multicolumn{1}{r}{(15.49\%)}&\multicolumn{1}{r}{(30.48\%)}\\
\bottomrule
\end{tabular}}\\[10pt]
\end{table}

\subsection{Computational kernel performance}\label{kernel_performance}

Table \ref{table:kernel_performance_viscous} shows the computational performance per CMG (1 process) of the viscous term kernel, which dominates 25\% of the cost of the time integration loop. The upper limit estimate was calculated based on the ratio of the amount of memory access to the number of operations (Byte per Flop) required from code and provided from the system. This kernel is optimized with various source modifications, such as loop collapse, memory access sequentialization, exchanging array dimension, loop fission, and loop blocking. Due to the optimization, the peak performance ratio has improved from the original 7.03\% to 13.95\%. Estimates suggest a further 2.7\% improvement.

\begin{table}[h]
\small\sf\centering
\caption{Viscosity term calculation performance measured value and upper limit estimated value (when running 2.0 GHz).\label{table:kernel_performance_viscous}}
\scalebox{0.95}{
\begin{tabular}{rrr}
\toprule
Original&Optimized&Upper limit\\
\midrule
53.87 GFLOPS&106.75 GFLOPS&128.07 GFLOPS\\
(7.03\%)&(13.95\%)&(16.67\%)\\
\bottomrule
\end{tabular}}\\[10pt]
\end{table}

Table \ref{table:kernel_performance_convec} shows the computational performance per CMG (1 process) of the high-cost convection term kernel, which dominates 50\% of the cost of the time integration loop. This kernel is not a memory throughput bottleneck because the ratio of the source memory access to the number of operations (Byte per Flop) is below the value the system can provide. Therefore, it is not possible to calculate the upper limit of the performance estimated value by Byte per Flop. Currently, this kernel applied some optimization (loop fission and array dimension exchange) indicates 14.7\% of the peak performance, but a significant performance improvement can be expected by computing performance optimization for utilizing software pipelining and SIMD.

\begin{table}[h]
\small\sf\centering
\caption{Convection term calculation performance measured value and upper limit estimated value (when running 2.2GHz)).\label{table:kernel_performance_convec}}
\scalebox{0.95}{
\begin{tabular}{rrr}
\toprule
Original&Optimized&Upper limit\\
\midrule
82.23 GFLOPS&123.98 GFLOPS&–\\
(10.10\%)&(14.70\%)&(–\%)\\
\bottomrule
\end{tabular}}\\[10pt]
\end{table}

\subsection{Weak scaling performance}\label{weak_scaling_performance}

\begin{figure}[H]
\centering
\includegraphics[width=0.3\textwidth]{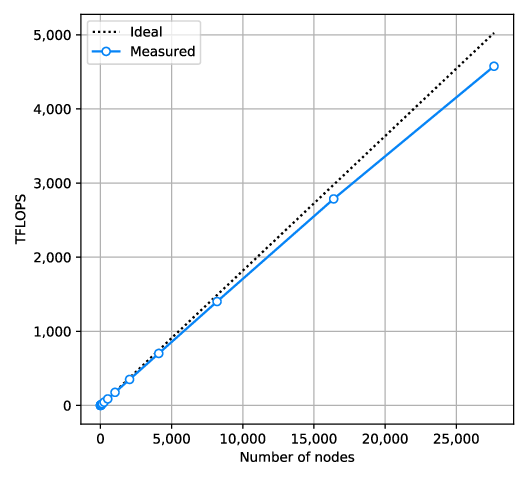}
\caption{Weak scaling results of cavity flow problem.\label{fig:weak_scaling_cube_cavity}}
\end{figure}

Weak scaling performance is first evaluated using the idealized cavity problem, which is grown along the y-axis direction proportional to the number of nodes, with communication conducted with the immediate neighbours regardless of the number of nodes. Although seemingly trivial, we have no collision in packets resulting in communication congestion due to the 6D torus topology of FUgaku, unlike typical machines with e.g. reduced Fattree network where congestion often occurs with scaling. In order to gurantee collision-free communication, we map the the processes responsible for the adjacent subdomains to be mapped to the neighboring nodes according to the TofuD topology.
Table \ref{table:weak_scaling_cube_cavity} and Figure \ref{fig:weak_scaling_cube_cavity} shows the results of weak scaling from 1 node to 27,648 nodes of Fugaku using the ideal problem; execution time is 27.24 seconds for 27,648 nodes compared to 25.48 seconds for one node, or 91.07\%, almost perfect weak scaling. Here the problem space consists of 21.7 billion cells, and the performance attained is 4.5 PFLOPS.

\begin{table}[h]
\small\sf\centering
\caption{Weak scaling results of cavity flow problem.\label{table:weak_scaling_cube_cavity}}
\scalebox{0.95}{
\begin{tabular}{rrrrr}
\toprule
\multicolumn{1}{l}{Nodes}&Cells&Exec. time&\multicolumn{1}{l}{TFLOPS}&\multicolumn{1}{l}{Scaling}\\
\midrule
1&786K&25.48 s&0.18 (5.38\%)&100.00\%\\
256&201M&26.94 s&44.00 (5.09\%)&94.57\%\\
512&402M&26.92 s&88.09 (5.09\%)&94.67\%\\
1,024&805M&26.86 s&176.55 (5.10\%)&94.87\%\\
4,096&3.2G&27.05 s&701.37 (5.07\%)&94.22\%\\
8,192&6.4G&27.07 s&1,401.68 (5.06\%)&94.15\%\\
16,384&12.8G&27.24 s&2,785.88 (5.03\%)&93.56\%\\
27,648&21.7G&27.98 s&4,575.74 (4.90\%)&91.07\%\\
\bottomrule
\end{tabular}}\\[10pt]
\end{table}

\begin{figure}
\centering
\includegraphics[width=0.3\textwidth]{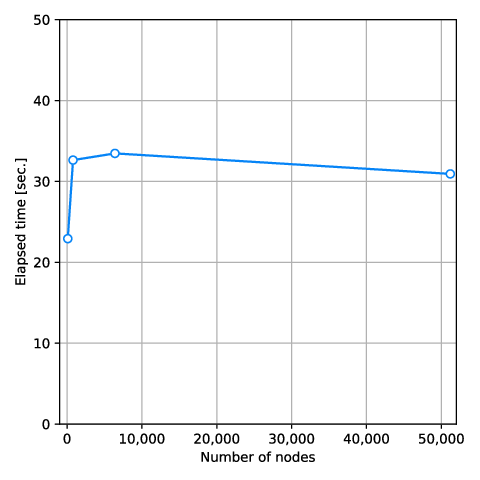}
\caption{Weak scaling results of vehicle model.\label{fig:weak_scaling_cube_vehicle}}
\end{figure}

\begin{figure}
\centering
\includegraphics[width=0.5\textwidth]{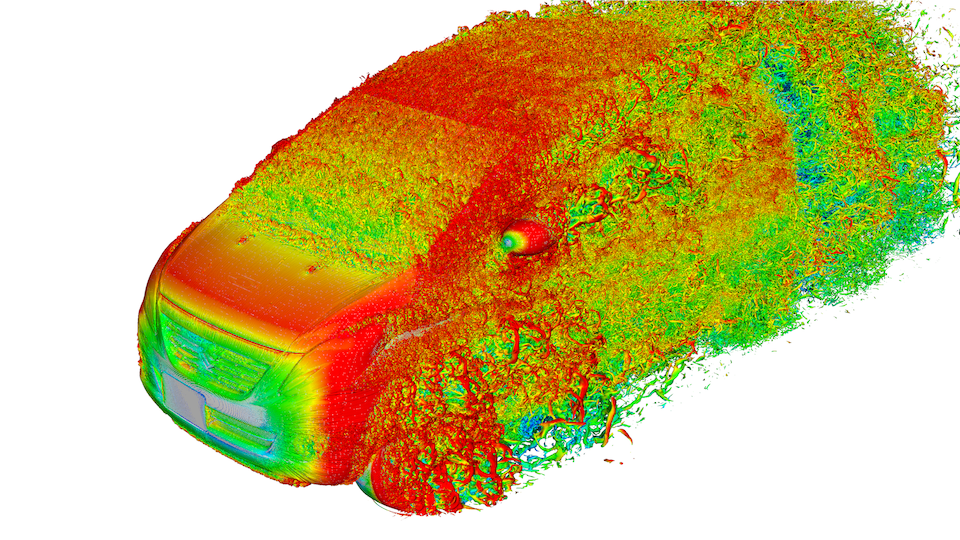}
\caption{Flow simulation around vehicle body (5.58 billion cells). \label{fig:vehicle}}
\end{figure}

We also measured the weak scaling of a real vehicle model (Figure \ref{fig:vehicle}).
A vehicle model (130 million cells) was used as a basis, and then the mesh was subdivided into two along all three axis step by step so that the number of cells is eight times larger, and then assigned to the nodes so that the number of grid points within the nodes remain the same. Compared to the ideal example, communication becomes much more complex, and load balancing becomes essential.
Table \ref{table:weak_scaling_cube_vehicle} and Figure \ref{fig:weak_scaling_cube_vehicle} show the result of evaluating the change in elapsed time when the number of nodes is increased for the time integration loop. Although being applied to industry-grade problem, scaling to 51,200 nodes from 100 nodes, which is the minimum number to fit the problem, was 74\%, excellent result for hierarchical grid-based solvers, and was essential for coping with COVID-19 problems requiring high-resolution simulations, as we will have the free choice of running more problems with higher throughput, or going higher resolution with high scalability, depending on the problem requirements.

\begin{table}[h]
\small\sf\centering
\caption{Weak scaling results of vehicle model.\label{table:weak_scaling_cube_vehicle}}
\scalebox{0.95}{
\begin{tabular}{rrrrr}
\toprule
\multicolumn{1}{l}{Nodes}&Cubes&Cells&\multicolumn{1}{l}{Time}&\multicolumn{1}{l}{Scaling}\\
\midrule
100&32K&132M&22.92 s&100.00\%\\
800&260K&1.0G&32.63 s&70.23\%\\
6,400&2.0M&8.5G&33.46 s&68.49\%\\
51,200&16.6M&68.2G&30.92 s&74.11\%\\
\bottomrule
\end{tabular}}\\[10pt]
\end{table}

\section{Fighting COVID-19 Droplet/Aerosol Transmissions with Fugaku over Past Year and a Half}

\subsection{How Fugaku was used in the Fight against COVID-19}

More important 'performance' was the throughput of the COVID-19 droplet/aerosol simulation we were able to conduct using substantial portion of Fugaku over a continuous period of more than a year and a half, and actually achieve digital transformation of epidemiology as well as the society with high-end HPC, not just performance of a singleton application run.

\begin{figure*}
\centering
\includegraphics[keepaspectratio,width=\textwidth]{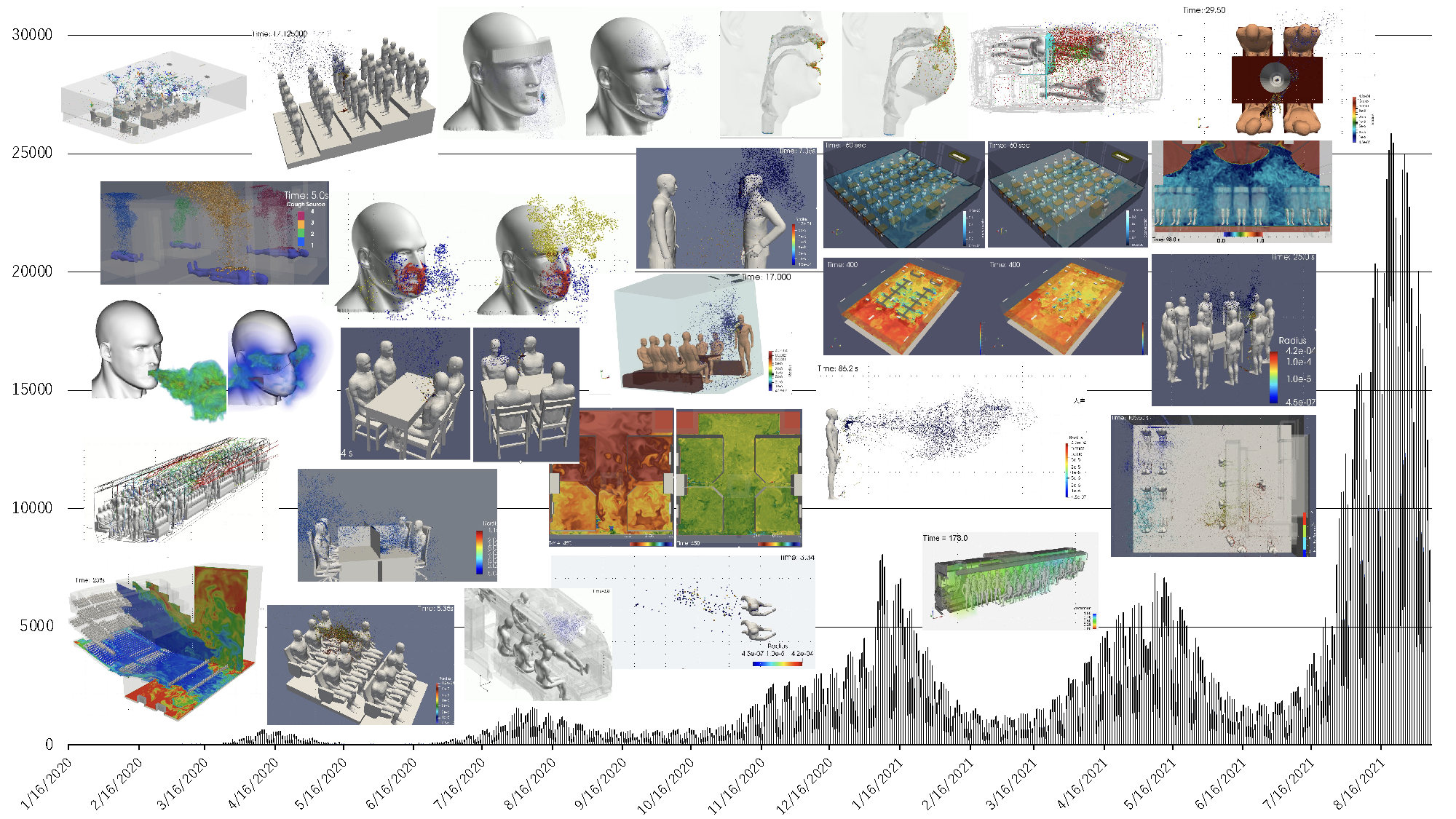}
\caption{Some of the visualization cases against the changes in the number of daily new infected people in Japan.\label{fig:simulation_cases}}
\end{figure*}

Our battle have been extensively conducted by formulating a comprehensive industry-academia-government collaboration led by RIKEN, as well as Kobe Univ., Toyohashi Univ. of Tech., Kyoto Institute of Tech., Osaka Univ., Tokyo Institute of Tech., Kyushu Univ., Kashima Corp., and Daikin Industries Ltd. as steering members. Toyota Motor Corp., Japan Airlines Co., Ltd., Daio Paper Corp (face masks)., Mitsubishi Fuso Truck and Bus Corp., Suntory Liquors Ltd., Toppan Inc., UNIQLO Co. Ltd. etc. participated providing actual analysis targets for their products as well as the resulting societal situations. Despite the rapid time-to-solution as described, there were limits to the amount of human resources on the academia side, and as such, the evaluation targets had to be carefully prioritized according to the infection risks as well as other societal situations of Japan; for example, early on in the pandemic, it was essential to determine whether masks were effective, as well as to assess the risks and come up with effective mitigation measures for commuter trains and office spaces, as Japan cannot be constitutionally locked down to prevent people to commute to their offices. The results of the simulations were visualized, and press conferences were held every 1--2 months, with usually around 100 media persons participating. Figure \ref{fig:simulation_cases} shows a sample of simulation cases we have made available through the press, along with changes in the daily number of new infected individuals in Japan. The simulation results in turn was broadcast to the Japanese public and overseas, with more than 350 TV/radio shows, 325 newspapers, and 1400 web articles, and contributed significantly to the social awareness of understanding the droplet/aerosol infection and importance of countermeasures.

In addition to disseminating information to the media, collaborations with government agencies such as MEXT, MLIT (Ministry of Land, Infrastructure, Transport and Tourism), and policy makes such as the Cabinet office were also actively conducted by providing the scientific data of infection risks for various societal situations of interest as identified by the ministries and their committees. The risk assessment data was continuously utilized to update the guidelines for resuming social and economic activities.
This research activities have been extensively conducted by formulating a comprehensive industry-academia-government collaboration led by RIKEN, as well as Kobe Univ., Toyohashi Univ. of Tech., Kyoto Institute of Tech., Osaka Univ., Tokyo Institute of Tech., Kyushu Univ., Kashima Corp., and Daikin Industries Ltd. as steering members, and Toyota Motor Corp., Japan Airlines Co., Ltd., Daio Paper Corp (face masks)., Mitsubishi Fuso Truck and Bus Corp., Suntory Liquors Ltd., Toppan Inc., UNIQLO Co. Ltd. etc. participated providing actual analysis targets for their products as well as the resulting societal situations. Despite the rapid time-to-solution as described, there were limits to the amount of human resources on the academia side, and as such, The evaluation targets had to be carefully prioritarized according to the infection risks as well as other societal situations of Japan; for example, early on in the pandemic, it was essential to determine whether masks were effective, as well as to assess the risks and come up with effective mitigation measure for commuter trains and office spaces, as Japan cannot be constitutionally locked down to prevent people to commute to their offices. The results of the simulations were visualized, and press conferences were held every 1--2 months, with usually around 100 media persons participating online from TV, newspapers, magazines, and Net media who immediately published the core results to the public. Figure \ref{fig:simulation_cases} shows a sample of simulation cases we have made available through the press, along with changes in the daily number of new infected individuals in Japan. The simulation results have been picked up by major media in Japan and overseas, with more than 350 TV/radio, more than 325 newspapers, and more than 1400 web articles, and contributed a lot to the social awareness of understanding the droplet/aerosol infection and importance of countermeasures.

\begin{figure}
\centering
\includegraphics[width=0.5\textwidth]{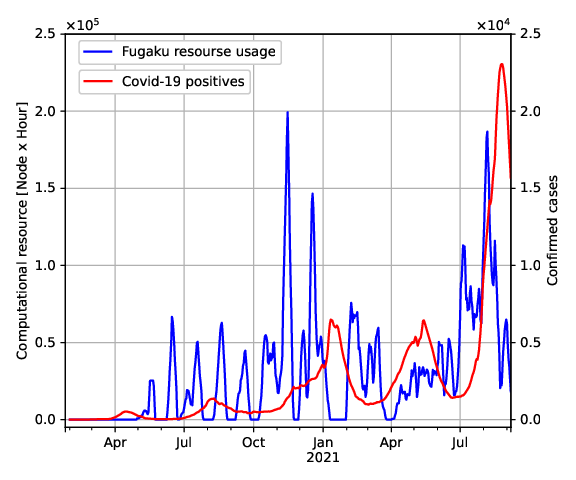}
\caption{Resource usage of Fugaku (left axis) and newly confirmed positives of Covid-19 in Japan (right axis).\label{fig:timeseries_moving_average}}
\end{figure}

Figure~\ref{fig:timeseries_moving_average} 
indicates the resource usage of Fugaku for this project in daily node hours, superimposed with the number of daily new infected people in Japan. It is interesting to note that there is a correlation between the peaks of weekly new positives and resource usage, and latter peak comes about one month earlier than the former. This is because the purpose of our activity is not to affect the transmissions when the infections are rising---this was better served by direct societal restrictions of certain degrees, such as closing down of shops and restaurants etc. Rather when the restrictions are relaxed for recovery of socio-economic activities, it is essential to make sure people avoid the unwarranted risks to for the recovery to be quick and sustainable. Hence, the announcement of new results was always timed around the early part of the recovery phase, identifying new high risk situations coinciding with the society re-opening.

 As of the end of September 2021, we have already consumed the computational resources of about 17.5 million node hours of Fugaku, which correspond to over 30 million node hours of machines facilitating  Intel® Xeon processors, and 40 million for Xeon Phi, according to our earlier benchmark. This is nearly matches the the capacity of exclusively dominating the OakForest-PACS machine at the Univ. of Tokyo with over 8000 Xeon Phi nodes and the former No.1 and currently the fifth fastest machine in Japan and 32nd globally according to the Top500 according to latest June 2021 ranking, for an entire year assuming about 80\% utilization.
 
 Overall, 1049 societal cases so far has been simulated over the course of 17 months or approximately 62 cases per month throughput, including geometry generation from CAD data or laser point cloud measurements, actual droplet/aerosol simulation, and post-processing including visualization. Since they are too numerous to describe, we will cover only three examples, but many more had transformational affect to the behavior of the entire Japanese public to curtail the pandemic.

\begin{table}[h]
\small\sf\centering
\caption{Total simulation cases, sizes, and computational costs.\label{table:cases}}
\scalebox{0.8}{
\begin{tabular}{p{3.5cm}p{7cm}rrrrr}
\toprule
Release Date&Targets&\multicolumn{1}{c}{Total}&Nodes&Node-hours&\multicolumn{1}{c}{Total}&\multicolumn{1}{c}{Total}\\
&&\multicolumn{1}{c}{mesh num.}&\multicolumn{1}{c}{/case}&\multicolumn{1}{c}{/case}&\multicolumn{1}{c}{cases}&\multicolumn{1}{c}{Node-hour}\\
\midrule
June, 17th, 2020  &Surgical masks (Gap)&29M&146&438&70&30,660\\
June, 17th, 2020  &Humidity and droplet transport on the table&34M&206&9,888&32&316,416\\
June, 17th, 2020  &Commuter train (windows opening)&163M&415&29,880&10&298,800\\
June, 17th, 2020  &Four-seated table (partitioning)&34M&206&9,888&32&316,416\\
June, 17th, 2020  &Office (windows opening etc.)&50M&511&24,528&10&245,280\\
June, 17th, 2020  &Hospital room (droplet dispersion)&48M&291&20,952&18&377,136\\
August, 24th, 2020&Surgical and fabric masks&29M&146&438&65&28,470\\
August, 24th, 2020&Face masks to protect against infection&35M&178&534&35&18,690\\
August, 24th, 2020&Hospital room (A/C effect)&48M&291&20,952&32&670,464\\
August, 24th, 2020&Office (Partitioning and A/C)&50M&511&24,528&10&245,280\\
August, 24th, 2020&Classroom (windows opening)&47M&330&7,920&16&126,720\\
August, 24th, 2020&Concert hall (ventilation)&135M&1,369&65,712&10&657,120\\
August, 24th, 2020&Concert hall (audience seats)&72M&440&21,120&32&675,840\\
October, 13th, 2020&Masks to protect against infection&35M&178&534&10&5,340\\
October, 13th, 2020&Speaking/singing/coughing&21M&158&7,584&24&182,016\\
October, 13th, 2020&Face shield&35M&213&10,224&16&163,584\\
October, 13th, 2020&Mouth shield&35M&213&10,224&16&163,584\\
October, 13th, 2020&Droplet transport (Humidity)&35M&206&9,888&24&237,312\\
October, 13th, 2020&Four-seated table (positioning)&26M&158&7,584&32&242,688\\
October, 13th, 2020&Chorus activity&40M&611&4,888&10&48,880\\
November, 26th, 2020&Face masks (difference of material)&29M&146&438&10&43,800\\
November, 26th, 2020&Karaoke room (positioning/partitioning)&50M&300&24,000&15&360,000\\
November, 26th, 2020&Outdoor activity (BBQ)&39M&239&11,472&24&275,328\\
November, 26th, 2020&Taxi&100M&480&11520&80&921,600\\
November, 26th, 2020&Commuter train (doors)&163M&415&29,880&5&149,400\\
November, 26th, 2020&Aircraft cabin&89M&546&39,312&20&786,240\\
November, 26th, 2020&Four-seated table (mouth shield)&35M&213&10,224&42&429,408\\
March, 4th, 2021&Face Masks (double wearing)&29M&75&225&15&3,375\\
March, 4th, 2021&Face Masks (nose fitter)&29M&75&225&15&3,375\\
March, 4th, 2021&Social distancing while walking&64M&391&18,768&28&525,504\\
March, 4th, 2021&Izakaya restaurant&87M&530&38,160&20&763,200\\
March, 4th, 2021&Ambulance&189M&740&17,760&55&976,800\\
June, 23th, 2021&Izakaya restaurant&87M&530&38,160&20&763,200\\
June, 23rd, 2021&Social distancing (variant)&21M&158&7,584&24&182,016\\
June, 23rd, 2021&Izakaya restaurant (risk map)&87M&530&38,160&8&305,280\\
June, 23rd, 2021&Four-seated table (portable ventilation)&20M&200&4,000&15&60,000\\
June, 23rd, 2021&Ceiling fan&132M&1,342&96,624&5&483,120\\
June, 23rd, 2021&Olympic stadium (audience seats)&83M&507&24,336&30&730,080\\
Ongoing&New Mouth guard&96M&585&28,080&24&673,920\\
Ongoing&Hospital (Sickroom)&120M&713&34,224&14&479,136\\
Ongoing&Hospital (Consultation room)&228M&1,393&100,296&5&501,480\\
Ongoing&Live Music Hall&137M&600&28,800&15&432,000\\
Ongoing&Live Music Hall (with Audience)&135M&600&28,800&15&432,000\\
Ongoing&Live Music Hall (droplet dispersion)&186M&960&46,080&8&368,640\\
Ongoing&Izakaya restaurant (risk map)&87M&530&38,160&8&305,280\\
Ongoing&Karaoke room (risk map)&80M&402&40,200&10&402,000\\
Ongoing&Museum (Planetarium)&117M&550&26,400&4&105,600\\
Ongoing&School gym (windows opening)&145M&482&11,568&4&46,272\\
Ongoing&Lecture room (A/C effect)&76M&568&13,632&4&54,528\\
Ongoing&Lecture room (droplet dispersion/coughing)&74M&568&40,896&9&368,064\\
Ongoing&Lecture room (droplet dispersion/speaking)&81M&568&40,896&4&163,584\\
\midrule
&&\multicolumn{3}{r}{Aggregate Total Cases \& Node Hours}&1,049&17,144,926\\
\bottomrule
\end{tabular}}\\[10pt]
\end{table}

\begin{figure*}
\centering
\includegraphics[keepaspectratio,width=\textwidth]{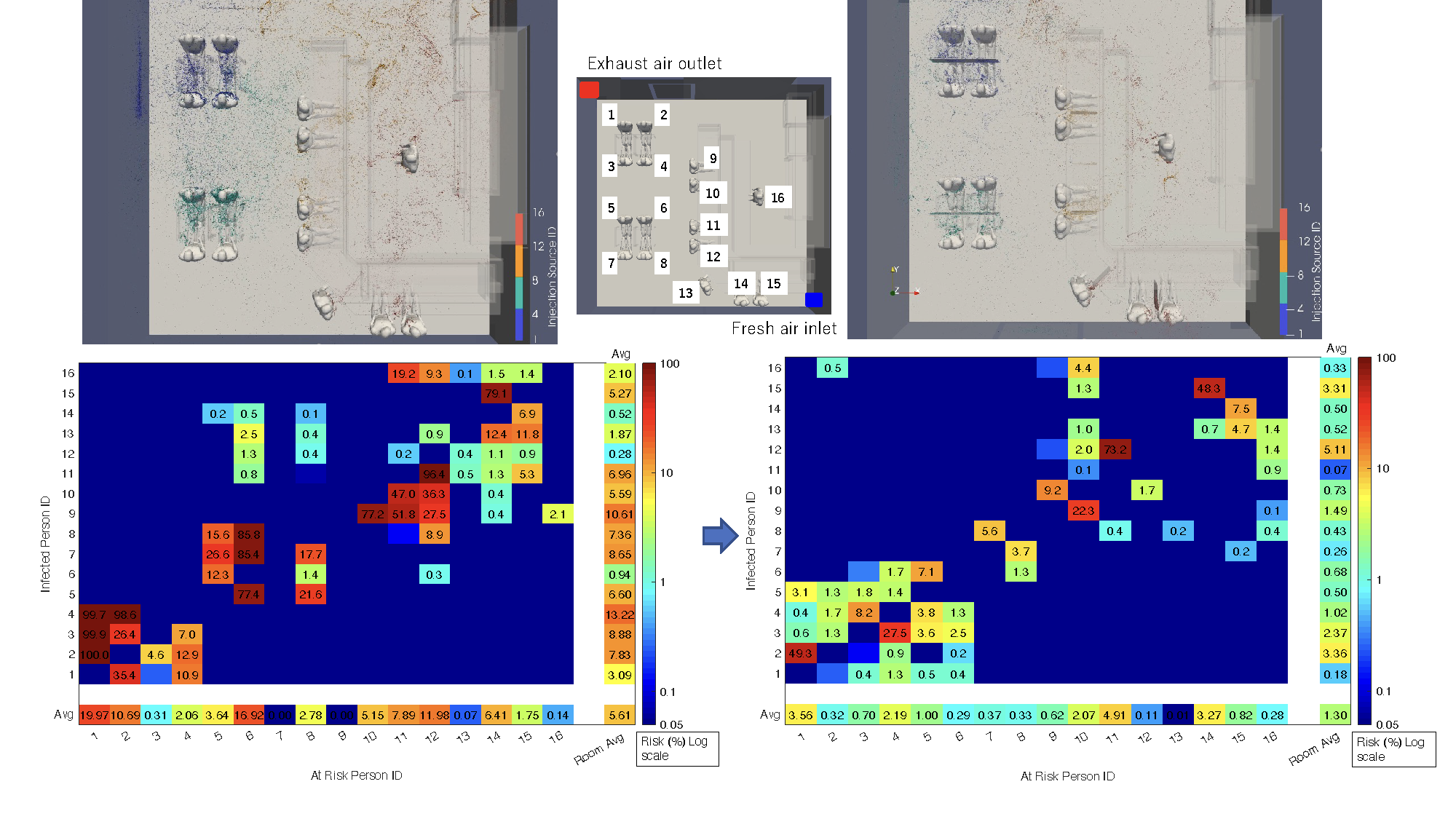}
\caption{Visualization of droplet/aerosol dispersion in a small restaurant (above) and the probability of droplet/aerosol infection (below) (left: default; right: A/C, kitchen duct, and partitioning).\label{fig:infection_risk}}
\end{figure*}

\subsection{Face Masks and Social Distancing}

The effect of face masks for reducing the infection risk and the evaluation of social distancing have attracted attention as social awareness since the early stage of our activities. These results include the analysis of mask materials---fabric masks were once popular in Japan when the unwoven throwaway masks were sold out, and even distributed proactively by the government in the fall of 2020; urethane masks were also popular among younger generations. Results from Fugaku, combined with physical experiments providing initial conditions, showed that such masks were significantly inferior to unwoven masks in containing the droplets and aerosols, and thus not recommended. 

By the time more infectious strains of virus emerged, we were able to come up with the first version of the end-to-end infection model, by combining the traditional epidemiological model in \ref{subsect:infection} with cumulative model of exhalation of aerosols versus their inhalation---we quantitatively assesses the infection risk by estimating the amount of virus taken into a human body through droplets/aerosols inhalation, after being exhaled in the room from the infected person. For simple cases, this becomes a function of distance and time, quantitatively characterizing 'social distancing'. There were many important results that were obtained and communicated to the public, such as the significant divergence of effectiveness of partition according to height which lead to redesigns throughout Japan, utmost importance of ventilation, relative {\em ineffectiveness} of double masking contrary to CDC recommendations, and the new delta variant requiring more than double the social distance of over 2 meters to attain the same probability as 1 meter which was sufficient for the original strain, etc. In all cases the results were visualized in detail, so that the general public 'saw' how they would be enveloped in infectious aerosols, tremendously raising awareness for the public to follow the safety protocols.

\subsection{Public Transportation}

Evaluating and controlling the risk of infection in public transport has been an important target since the early days of COVID-19 pandemic in order to prevent the spread of infection and balance socio-economic activities. So far, we have analyzed commuter trains, aircraft, commuter buses, sightseeing buses, and taxis, in addition to the ambulances at the request of Kobe City to reduce the risk for the paramedics. All simulations included detailed geometries of the vehicles, including the interior features such as chairs and handles as well as passengers taking various poses, as well as countermeasures such as shields in taxis.

In general, we have found public transportation to be relatively safe, and can be made safer by small countermeasures. For commuter trains, in addition to mechanical ventilation, we performed quantitative evaluations of the additional increase in ventilation efficiency by opening windows as well as the effect of the doors being opened at stations, and found both to be very effective to achieve very good ventilation for significant risk reductions. For taxis, in collaboration with Toyota, we modeled the airflow of the whole vehicle including the engine bay; here we found that contrary to intuitive belief, opening the passenger windows were not as effective as turning the mechanical ventilation with outside air being let in. This is primarily due to low speed of taxis at city speeds, combined with the relative smallness of the cabin space compared to trains, wherein the pressure difference between fore and aft would not be sufficiently cause circulatory air flow to allow for effective exchange of air with outside. After this finding, many taxi companies changed their anti-COVID protocols to turn on external air ventilation instead of opening the windows.

\subsection{Public Facilities}

Infection risk in public facilities has been evaluated for many sites, at offices, hospital rooms, elementary school classrooms, university lecture rooms, museums, live music club, concert halls, and sport stadiums. Initially, our results were limited to quantitative assessment and visualization of the droplets/aerosols, but the end-to-end model now allows the infection risk for the entire facility to be assessed in detail. Figure \ref{fig:infection_risk} shows the evaluation of the droplet/aerosol infection risk in a small restaurant with a capacity of 16 people, and effect of countermeasures to reduce the risk. The simulation calculates the expectation value of infection as a risk factor, when one infected person would be seated in one of the indexed places from 1 to 16 with equal probability, and everyone in the restaurant would be making conversation during their stay, say for one hour in this shown example. Vertical index indicate where the infected person would be seated, and horizontal index indicate the transmission risk of the person seated there by the infected person in the room. The overall expectation value of infection in this restaurant would be 5.6\%, which may seem relatively small, but if the restaurant would be open for 10 hours, we would observe 1 transmission per every two days, and much more with new strains . Once air conditioning and kitchen duct are operated to enhance the ventilation and diffusion of  air together with the partitions placed on the tables and the counter, the risk would be reduced to 1.3\%. By Utilizing such simulation with scientifically trustworthy quantitative evaluations of countermeasures in various public facilities and settings, the Japanese policy makers were able to propose effective infection control measures that balance infection prevention and economic activities, without mandated lockdowns as have been observed in most other countries. Such forced measures are considered unconstitutional in Japan. So, all societal restrictions had been basically voluntary; but still, the public largely followed the government's anti-COVID request and recommendations, as they had been presented as being simulated on Fugaku, which had won the public trust in providing scientific evidences.

\section*{Acknowledgement}

Computational resources of Fugaku were provided through the HPCI System Research Project (Project ID: hp210086, hp210242, hp210262) as well as the Fugaku's COVID-19 project by MEXT with Riken. This work has also been partially supported by JST CREST Grant Number JPMJCR20H7, Japan. We sincerely thank Prof. Akiyoshi Iida (Toyohashi University of Technology), Prof. Masashi Yamakawa (Kyoto Institute of Technology), Prof. Naoki Kagi (Tokyo Institute of Technology), and Prof. Kazuhide Ito (Kyushu University)  for their invaluable advice, support and discussions relating to this work. We are also grateful to Kajima Corporation and Daikin Industries, Ltd. for collaborative support from the early stages of this work.

\bibliographystyle{plain}
\bibliography{main}

\end{document}